\documentclass[iop]{emulateapj}

\shorttitle{A Successful Broad-band Survey for Giant \lya\ Nebulae I}
\shortauthors{Prescott et al.}

\usepackage{longtable}

\newcommand{\bw}{$B_{W}$}
\newcommand{\lya}{\hbox{{\rm Ly}\kern 0.1em$\alpha$}}
\newcommand{\oii}{[\hbox{{\rm O}\kern 0.1em{\sc ii}}]}
\newcommand{\oiii}{[\hbox{{\rm O}\kern 0.1em{\sc iii}}]}
\newcommand{\halpha}{\hbox{{\rm H}\kern 0.1em$\alpha$}}
\newcommand{\heii}{\hbox{{\rm He}\kern 0.1em{\sc ii}}}

\newcommand{\numcand}{79} 
\newcommand{\numfirst}{39} 
\newcommand{\numsecond}{40} 
\newcommand{\numthird}{6} 
\newcommand{\numfirstthree}{32} 
\newcommand{\numfirstfour}{7} 
\newcommand{\numsecondthree}{31} 
\newcommand{\numsecondfour}{9} 
\newcommand{\numthirdthree}{1} 
\newcommand{\numthirdfour}{5} 
\newcommand{\numzero}{738} 
\newcommand{\numone}{491} 
\newcommand{\numtwo}{170} 
\newcommand{\numthree}{459} 
\newcommand{\numfour}{408} 
\newcommand{\numzerotwo}{1399} 
\newcommand{\numthreefour}{867} 
\newcommand{\numsources}{2266} 
\newcommand{\bootesarea}{9.4} 
\newcommand{\searcharea}{8.5} 
\newcommand{\rejexparea}{0.28} 
\newcommand{\rejimarea}{0.68} 
\newcommand{\sblimonesig}{28.9} 

\begin{document}

\title{A Successful Broad-band Survey for Giant \lya\ Nebulae I: Survey Design and Candidate Selection}

\author{Moire K. M. Prescott\altaffilmark{1,2}, Arjun Dey\altaffilmark{3}, Buell T. Jannuzi\altaffilmark{3}}

\altaffiltext{1}{TABASGO Postdoctoral Fellow; Department of Physics, Broida Hall, Mail Code 9530, University of California, Santa Barbara, CA 93106, USA; mkpresco@physics.ucsb.edu}
\altaffiltext{2}{Steward Observatory, University of Arizona, 933 N. Cherry Avenue, Tucson, AZ 85721, USA}
\altaffiltext{3}{National Optical Astronomy Observatory, 950 North Cherry Avenue, Tucson, AZ 85719, USA} 

\begin{abstract}
Giant \lya\ nebulae (or \lya\ ``blobs") are likely sites of ongoing massive galaxy
formation, but the rarity of these powerful sources has made it difficult to form a
coherent picture of their properties, ionization mechanisms, and space density.
Systematic narrow-band \lya\ nebula surveys are ongoing, but 
the small redshift range covered and the observational expense limit the comoving volume that 
can be probed by even the largest of these surveys and pose a significant problem when 
searching for such rare sources.  
We have developed a systematic search technique designed to find large \lya\ nebulae at $2\lesssim z \lesssim3$
within deep {\it broad-band} imaging and have carried out a survey of the \bootesarea\ 
square degree NOAO Deep Wide-Field Survey (NDWFS) Bo\"otes field.  With a total 
survey comoving volume of $\approx$10$^{8}$ h$^{-3}_{70}$ Mpc$^{3}$, this is the largest 
volume survey for \lya\ nebulae ever undertaken.  
In this first paper in the series, we present the details of 
the survey design and a systematically-selected sample of \numcand\ candidates, 
which includes one previously discovered \lya\ nebula.  
\end{abstract}


\section{Introduction}
\label{sec:intro}

Giant radio-quiet \lya\ nebulae \citep[e.g.,][]{francis96,stei00,pal04,mat04,dey05,sai06,nil06,smi07,greve07,yang09} 
likely provide an observational window into the physics of ongoing massive galaxy formation.  
As large as $\sim$100~kpc across, \lya\ nebulae emit copious \lya\ emission 
($\sim10^{44}$ erg s$^{-1}$), signaling the presence of highly energetic phenomena.  
\lya\ nebulae are sometimes surrounded by or associated with young, star-forming galaxy populations 
and obscured AGN \citep[e.g.,][Prescott et al. 2012c, submitted]{chap01,basu04,mat04,
dey05,gea07,gea09} and have been shown to reside in overdense environments \citep{mat04,mat05,mat09,sai06,pres08,yang09,yang10}.
Studying \lya\ nebulae can provide important insight into the process of  
galaxy formation, but after a decade of study, we have not yet developed a complete understanding 
of the dominant power source or, even more fundamental, the space density of \lya\ nebulae.

The problem lies in having such a small sample - only a few dozen large
($\sim100$~kpc) \lya\ nebulae are currently known - and in attempting to measure 
the space density of a rare population over restricted volumes.
Radio-quiet \lya\ nebulae were first discovered in targeted narrow-band imaging studies of known
overdensities \citep[e.g.,][]{francis96,stei00,mat04}.  
Since then, the sample has slowly grown as a result of ever more expensive 
narrow-band surveys over increasingly wide areas \citep[e.g.,][]{nil06,smi07,sai06,yang09,yang10,mat11}.  
Building a sufficient sample of such rare objects over a wide range of environments 
and accurately measuring their space density requires surveying larger comoving volumes
than are typically feasible via standard narrow-band imaging.

We have therefore taken a complementary approach by developing a systematic search algorithm 
to find luminous \lya\ nebulae at $2\lesssim z \lesssim3$ using deep {\it broad-band} imaging 
and have tested the technique using archival imaging data from the NOAO Deep 
Wide-Field Survey \citep[NDWFS;][]{jan99}.  
We describe our systematic broad-band \lya\ nebula survey in three papers.  
This paper (Paper~I) discusses the survey design and candidate selection approach.  
In Section~\ref{sec:design}, we describe the algorithm used to select 
\lya\ nebulae.  Section~\ref{sec:candidate} presents the sample of \numfirst\ first priority and \numsecond\ 
second priority \lya\ nebula candidates selected from archival imaging of the 
Bo\"otes Field of NDWFS and discusses the potential sources of contamination.  
In Section~\ref{sec:discussion}, we address the complementarity of our approach relative to 
traditional narrow-band surveys, and we conclude in Section \ref{sec:conclusions}.  
Discussions of (a) our successful spectroscopic follow-up campaign 
and (b) the survey selection function and implied \lya\ nebula number density 
are presented in the two subsequent papers in this series (Prescott et al. 2012b and 
Prescott et al. 2013, in preparation; hereafter, Paper~II and III).  

We assume the standard $\Lambda$CDM cosmology ($\Omega_{M}=0.3$, $\Omega_{\Lambda}=0.7$, $h=0.7$); 
1\arcsec\ corresponds to a physical scales of 8.5-7.6~kpc for redshifts of $z\approx2-3$.  All magnitudes
are in the AB system.


\section{Survey Design}
\label{sec:design}

While most of the early examples of the \lya\ nebula class were found 
via narrow-band surveys of known galaxy overdensities, one of the largest 
\lya\ nebulae was discovered in a very different manner \citep[LABd05;][]{dey05}.  
This \lya\ nebula at $z\approx2.7$ came to light during a study of strong 24$\mu$m sources detected 
by the Spitzer Space Telescope in the NDWFS Bo\"otes field.  
Strikingly, this source had a diffuse, extended morphology and very 
blue colors in deep broad-band imaging.  The authors suspected and 
spectroscopically confirmed that \lya\ line and blue continuum  emission were 
dominating the broad-band flux and that the diffuse morphology indicated the 
presence of a giant, spatially extended \lya\ nebula.  

Inspired by this discovery, we developed a search for \lya\ nebulae 
designed to work on deep broad-band imaging and applied it to the $\approx$9 square degree NDWFS Bo\"otes field dataset.  
Using a deep, wide-area, ground-based imaging survey like NDWFS allows us to probe an enormous 
comoving volume with existing data and significantly reduce the amount of new telescope time required 
to complete the program.  In that sense, our survey is complementary to the more sensitive but smaller volume 
narrow-band \lya\ nebulae surveys \citep[e.g.,][]{mat04,sai06,smi07,yang09,yang10,mat11}.  
In this section we describe the NDWFS dataset and introduce the search algorithm 
used to select a sample of \lya\ nebulae candidates.  

\subsection{NDWFS Broad-band Data}

Our survey technique was built and tested using the broad-band optical 
imaging from NDWFS, data which are available through the NOAO Science 
Archive.\footnote{NOAO Science Archive: http://archive.noao.edu/nsa/}  
The \bw\ and $R$-band images of the \bootesarea\ square degree Bo\"otes field were 
obtained using the Mayall 4m Telescope and the MOSAIC 1 prime focus camera.  
The Bo\"otes field was originally selected as a field suitable for a deep extragalactic 
survey because of its low 100$\mu$m background and N(H\textsc{i}) column density along the 
line of sight, and for being well positioned from Kitt Peak National Observatory 
during a period of the year with historically good observing conditions.  
The entire Bo\"otes field was covered using stacked observations of 27 overlapping MOSAIC 
pointings.  The resulting median seeing was 1.10\arcsec\ and 1.10\arcsec\ in the \bw\ and $R$ bands, respectively, 
and the median point-source depths were $\approx$26.3 and 25.2 AB mag (5$\sigma$; 2\arcsec\ diameter apertures).  
The median 1$\sigma$ \bw\ surface brightness limit is \sblimonesig\ mag arcsec$^{-2}$ (1.1\arcsec\ diameter apertures), 
which corresponds (in the case of pure line emission) to a line flux surface brightness 
limit of $2.48\times10^{-17}$ erg s$^{-1}$ cm$^{-2}$ arcsec$^{-2}$.  
The data were reduced by the NDWFS team using IRAF\footnote{NDWFS Data Processing: http://www.noao.edu/noao/noaodeep/}.

\subsection{The Search Algorithm}
\label{sec:algorithm}

\begin{figure}
\center
\includegraphics[angle=0,width=4in]{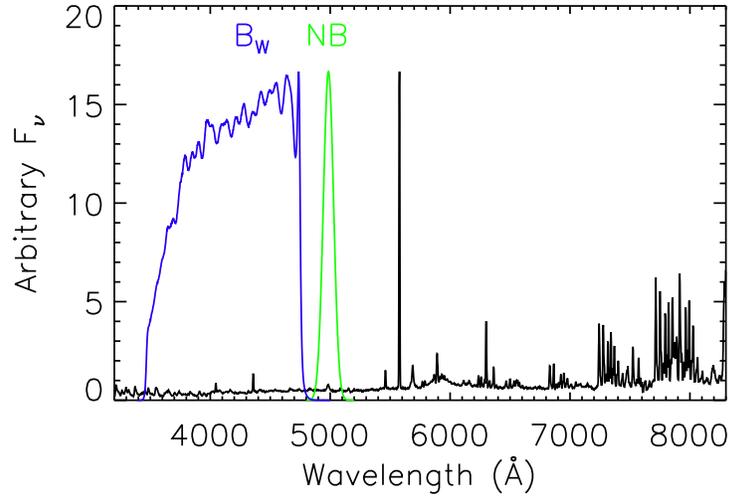}
\caption[Systematic Broad-band Search for Finding \lya\ Nebulae.]
{Broad-band \bw\ bandpass and a generic narrow-band filter bandpass (NB; FWHM~$=100$\AA) overplotted on a spectrum of 
the night sky.  Our systematic broad-band search technique, designed to find luminous \lya\ nebulae at $2\lesssim z \lesssim3$, 
is able to efficiently survey enormous comoving volumes by relying on deep, wide-field, broad-band imaging and the very dark sky in the blue.
}
\label{fig:bandpass}
\end{figure}

The goal of our systematic survey was to find large, luminous \lya\ nebulae similar to LABd05.  
The search pipeline uses a blue broad-band image --- in the case of the NDWFS, the 
\bw-band --- as the primary search image and is tuned to find diffuse, spatially extended, low surface 
brightness objects.  It is most sensitive to extended sources for which a bright \lya\ emission line boosts the 
broad-band flux relative to the very dark sky in the blue (Figure~\ref{fig:bandpass}), i.e.,  
the largest and brightest \lya\ nebulae.  A redder band --- the $R$-band in the case of NDWFS --- 
is used to derive secondary color information. 
 
\begin{figure}
\center
\includegraphics[angle=0,width=6.5in]{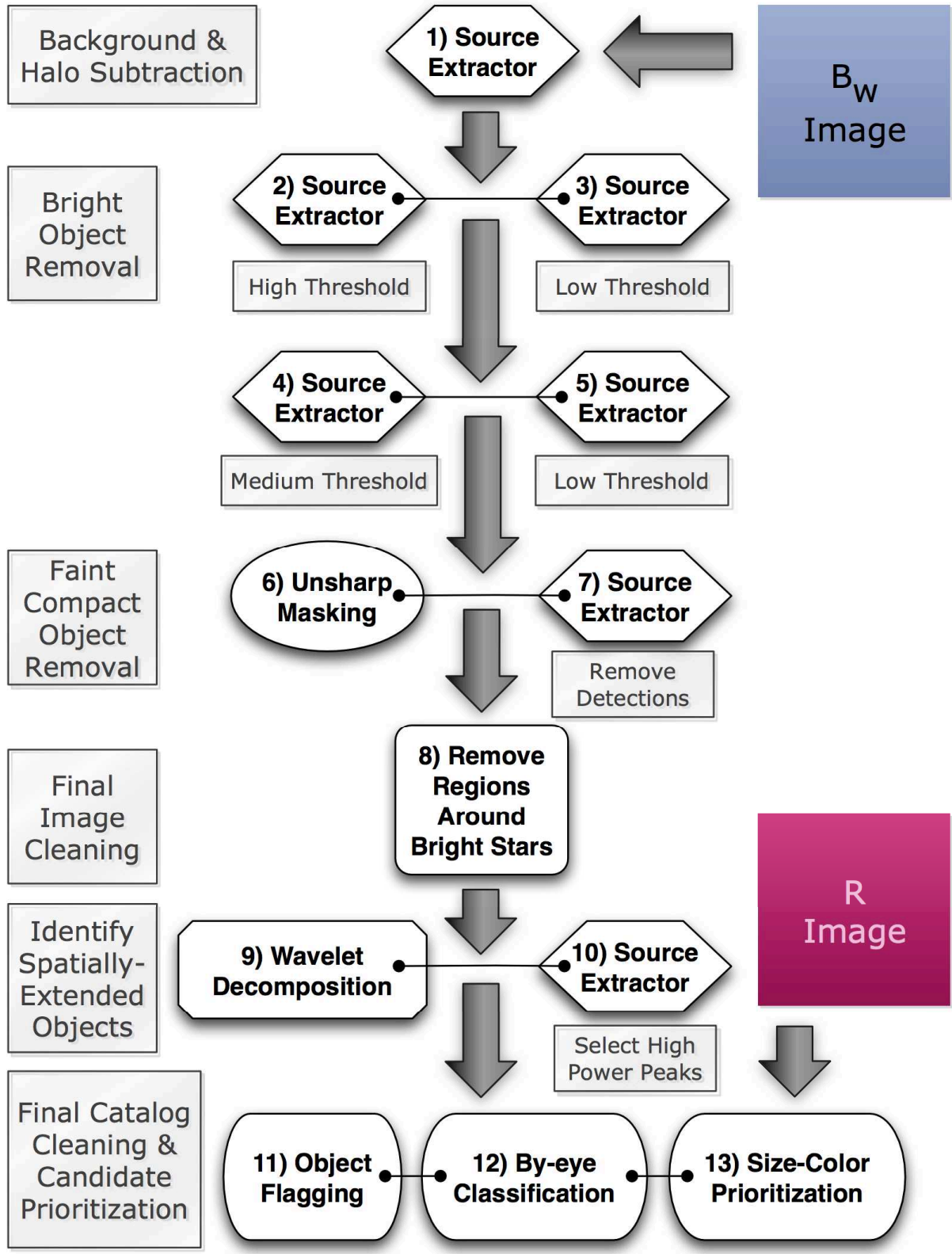}
\caption[Systematic Broad-band Search for Finding \lya\ Nebulae.]
{Flowchart outlining the individual steps of our systematic broad-band search for luminous \lya\ nebulae.
}
\label{fig:flowchart}
\end{figure}

\begin{figure}
\center
\includegraphics[angle=0,width=6.5in]{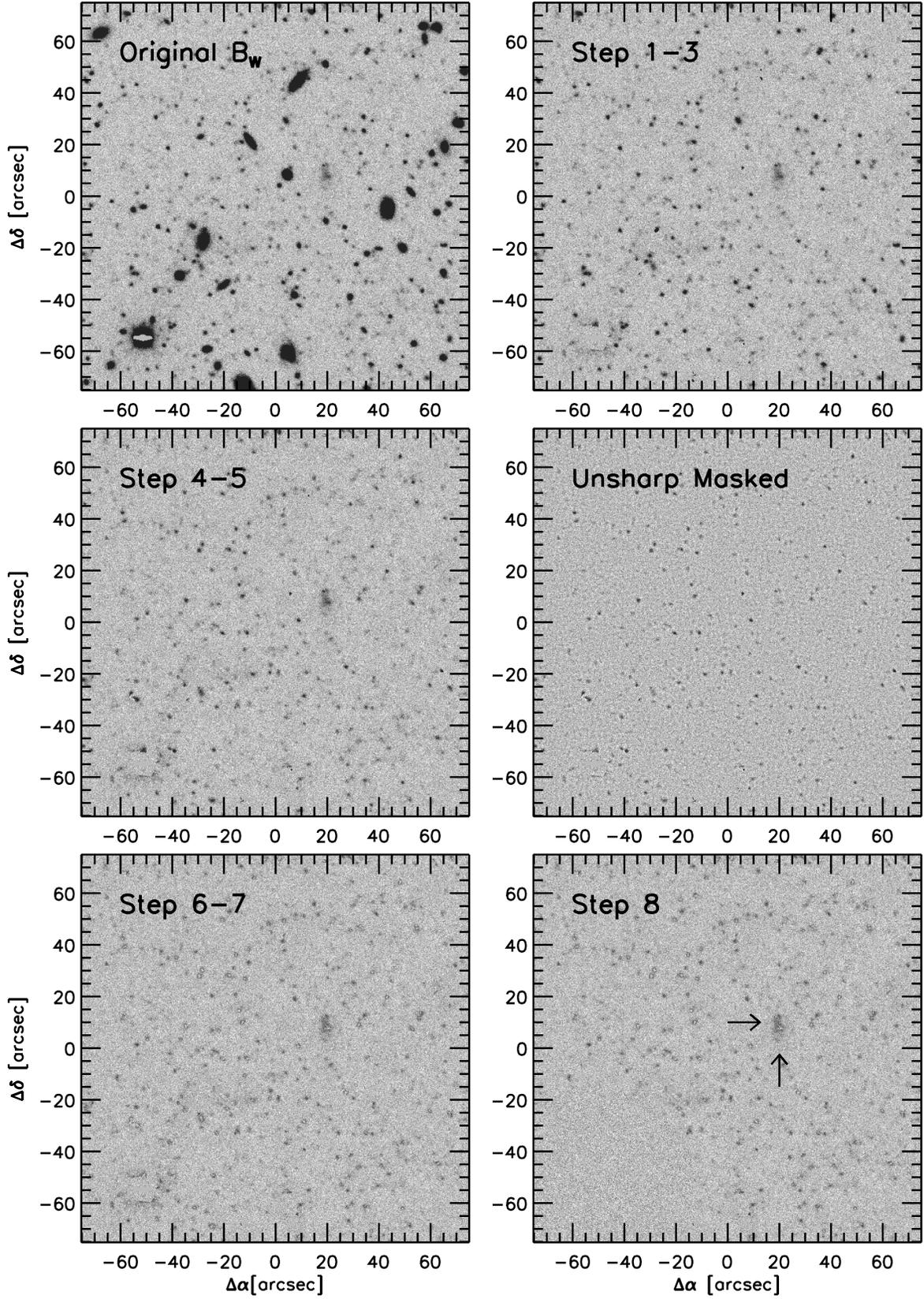}
\caption[Systematic Broad-band Search for Finding \lya\ Nebulae.]
{Individual steps of the broad-band \lya\ nebula search algorithm illustrated using the region 
around LABd05 (Section~\ref{sec:algorithm}).  
The diffuse emission of LABd05 is clearly visible (arrows) in the final frame ({\it Step 8}).
}
\label{fig:search}
\end{figure}

\begin{figure}
\center
\includegraphics[angle=0,width=6.5in]{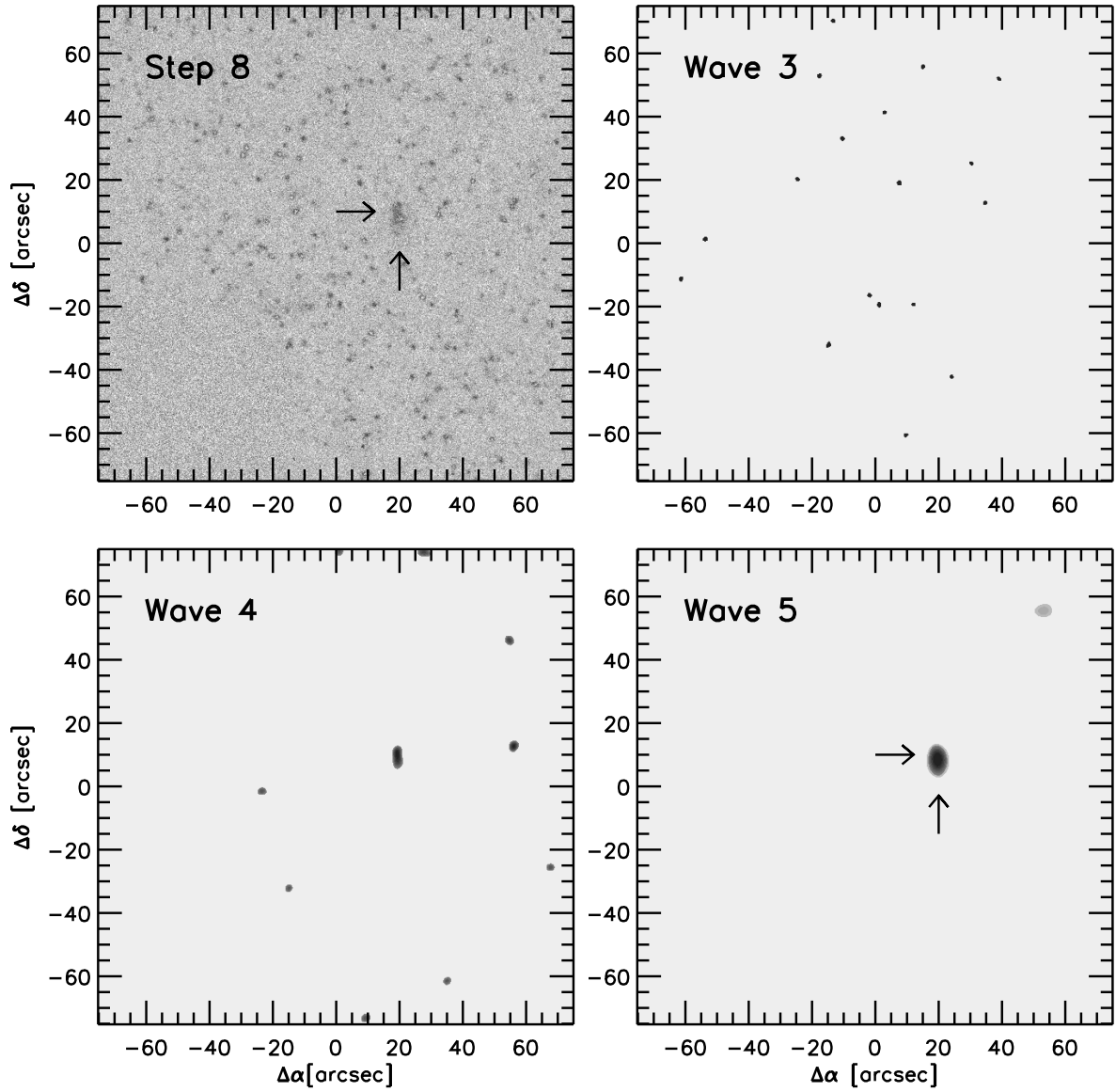}
\caption[Wavelet Decomposition Step of the Broad-band \lya\ Nebula Search.]
{Wavelet decomposition is used to separate out sources of different size scales.  
The upper left panel shows the final result of Steps 1-8 of the search algorithm, as in Figure~\ref{fig:search}.  
The other three panels show the wavelet power maps 
for different size scales.  The diffuse emission of LABd05 
generates a large wavelet power peak in the final panel and is easily 
selected as a \lya\ nebula candidate (arrows).  The source in the upper righthand corner 
of the final panel was below the 4$\sigma$ threshold (Section~\ref{sec:spatial}) and was not selected.
}
\label{fig:wave}
\end{figure}

Given the depth and the size of the imaging survey, in designing our search algorithm 
it was important to minimize the number of contaminant objects while retaining objects likely to be \lya\ nebulae.  
Our search strategy therefore involved first subtracting off bright galaxies and compact objects and then 
using a wavelet deconvolution algorithm to select diffuse, spatially extended, low surface brightness objects with blue colors.  
The search pipeline steps are described in detail below and illustrated using a flowchart\footnote{Flowchart created using 
Omnigraffle: http://www.omnigroup.com/products/omnigraffle/} in Figure~\ref{fig:flowchart}.  
Figures~\ref{fig:search} and \ref{fig:wave} show the pipeline at work on the image region near LABd05.  

\subsubsection{Background and Halo Subtraction}

{\it Step 1:}  The first step is to remove the sky background 
using a background map generated by Source Extractor \citep[Version 2.4.4;][]{ber96}.  
We used a high detection threshold ({\sc detect\_thresh}~$=$10$\sigma$ per pixel), set background parameters 
({\sc back\_size}~$=$64 and {\sc back\_filtersize}~$=$3), and required a minimum of 5 connected 
pixels ({\sc detect\_minarea}~$=$5).  This background map, which includes a contribution from the halos around 
bright stars, was subtracted from the search image.  

\subsubsection{Bright Object Removal}

In order to remove bright stars and galaxies that may contaminate the final candidate list, we employed the following procedure.  

{\it Step 2:}  
We first ran Source Extractor with a high detection threshold ({\sc detect\_thresh}$=$10$\sigma$ per pixel) 
and a minimum of 5 connected pixels ({\sc detect\_minarea}$=$5) in order to generate a catalog of 
bright sources.  

{\it Step 3:}  We then ran Source Extractor with a very low threshold ({\sc detect\_thresh}$=1\sigma$ per pixel) 
in order to generate a deep segmentation map, i.e., a map of all object pixels, that would 
include pixels in the faint wings of bright stars and galaxies.  This iteration used 
the {\sc assoc} mode in which the catalog from {\it Step 2} was used as an input to Source Extractor and 
only those detections that match were included in the catalog.  Detected 
sources were matched against the input catalog in a weighted fashion based on measured pixel position 
({\sc x\_image}, {\sc y\_image}), the object flux ({\sc flux\_auto}), and a search radius ({\sc assoc\_radius}$=$2 pixels).  
Using the deep segmentation maps from this step and the image rms maps, we replaced regions identified with 
bright stars and galaxies with patches of sky noise.  

{\it Step 4:}  We ran Source Extractor on the ``cleaned" image from {\it Step 3} 
but this time with an intermediate threshold ({\sc detect\_thresh}$=$6$\sigma$ per pixel, determined based on visual inspection) 
to further remove stars and galaxies.  

{\it Step 5:}  We ran Source Extractor on the image from {\it Step 4} with the {\sc assoc} feature in order to 
generate the corresponding deep segmentation map.  Again, the detections are replaced with sky noise.

\subsubsection{Faint Compact Object Removal}

{\it Step 6:}  We next used unsharp masking to remove fainter objects with compact morphology.  
An unsharp masked image was produced by smoothing the image from {\it Step 5} using 
an $11\times11$ pixel kernel and then subtracting the smoothed image from the original {\it Step 5} 
image.  In this process, diffuse emission was removed, leaving only compact sources 
in the unsharp masked image.  

{\it Step 7:}  We performed another {\sc assoc} Source Extractor run on the unsharp masked image 
in which we only allowed matched detections to be included in the catalog, i.e., anything detected in 
both the {\it Step 5} image and the unsharp masked image was reported.  
We removed these object regions by replacing them with sky noise, 
and in this way removed additional faint compact sources from the image.  

\subsubsection{Final Image Cleaning}

{\it Step 8:}  To avoid artifacts (e.g., saturated columns, poorly-subtracted halos) 
from bright stars being selected as sources later in the pipeline, 
we replaced the regions around bright stars ($B<16$ mag in the USNO-A2.0 catalog) 
out to a radius of 150 pixels ($\approx39$\arcsec) with sky noise.  This step 
removed $\approx$\rejimarea\ square degrees from the final survey area. 

\subsubsection{Identification of Spatially Extended Objects}
\label{sec:spatial}

{\it Step 9:}  We then applied a wavelet decomposition algorithm to decompose the image 
into maps showing the power on a number of different spatial scales.  

The basic idea behind wavelet decomposition is to decompose the image into
a set of ``wavelet planes,'' each of which contains the flux of sources
with power on that scale and the sum of which is the original image (ignoring losses).
The power of the wavelet decomposition approach is to reduce the interference
of small scale objects when looking for large scale structures.

We use the wavelet decomposition code {\it wvdecomp}\footnote{{\it wvdecomp} was written by 
Alexey Vikhlinin; 
{\sc zhtools} documentation and source code: http://hea-www.harvard.edu/RD/zhtools/.} 
to filter the image across six size scales ($scalemin=1$, $scalemax=6$, $threshold=3$, 
$thresholdmin=2$, $iter=10$).  
Here we give a brief description of the algorithm; a
detailed discussion of the code is given in the {\sc zhtools} documentation.
The {\it wvdecomp} algorithm uses an {\it \`a trous} wavelet kernel which is
defined as the difference between two functions ($f_{i}$ and $f_{i+1}$)
that each roughly resemble a Gaussian of width $2^{i-1}$.  On each scale,
the kernel is sensitive to objects of the same size ($\approx2^{i-1}$).  

The algorithm begins by convolving the image with the smallest wavelet kernel.
Any features that are insignificant in the convolved image are zeroed and
the result is subtracted from the original image, causing a majority of the flux
from small features comparable in size to the smallest wavelet kernel to be 
removed.  The process is repeated with larger and larger
wavelet kernels, and the results from each step are saved as wavelet power
maps or ``wavelet planes.''
To minimize the loss of information at each step, two thresholds are specified
in running the code: a detection threshold, t$_{max}$, and a filtering threshold,
t$_{min}$, both defined in terms of the image noise.
A source detected on a given scale above t$_{min}$ is counted as
significant only if the source local maximum also exceeds t$_{max}$.

{\it Step 10:} We used the known \lya\ nebula LABd05 to empirically determine the appropriate scale for large \lya\ nebulae.  
Using one final run of SourceExtractor, 
we generate a catalog of sources in the wavelet power map and select those with wavelet power peaks 
above 4$\sigma$, chosen to minimize the number of total candidates while preserving sources like 
LABd05.

\subsubsection{Final Catalog Cleaning and Candidate Prioritization}
\label{sec:finalcat}

{\it Step 11:}  Sources drawn from low exposure time regions of the survey ($<$4000~sec) 
as well as those found in regions that had been flagged 
in previous steps were removed from the candidate list.   
This step removed $\approx$\rejexparea\ square degrees from the final survey area. 

{\it Step 12:}  Postage stamps of candidates were inspected visually and classified by eye 
using morphological flags, where {\sc star} denotes a star halo, {\sc galaxy} denotes a galaxy halo, 
{\sc tidal/arm} indicates tidal tails or spiral arms, {\sc group} corresponds to a tight grouping of 
compact sources selected as a single large source, and {\sc diffuse} indicates 
spatially extended diffuse emission.  
The \numsources\ sources selected by the morphological search pipeline were categorized 
as follows: \numzero\ {\sc star} sources, \numone\ {\sc galaxy} sources, \numtwo\ {\sc tidal/arm} 
sources, \numthree\ {\sc group} sources, and \numfour\ {\sc diffuse} sources.  We removed 
those objects categorized as {\sc star}, {\sc galaxy}, and {\sc tidal/arm} from the candidate sample 
(\numzerotwo\ sources), since these are easy to classify unambiguously by eye.  
As the distinction between categories {\sc group} and {\sc diffuse} is more subjective, 
both category {\sc group} and {\sc diffuse} objects were included in the 
final sample (\numthreefour\ sources).  

\subsubsection{Candidate Selection}
\label{sec:selection}

\begin{figure}
\center
\includegraphics[angle=0,width=4in]{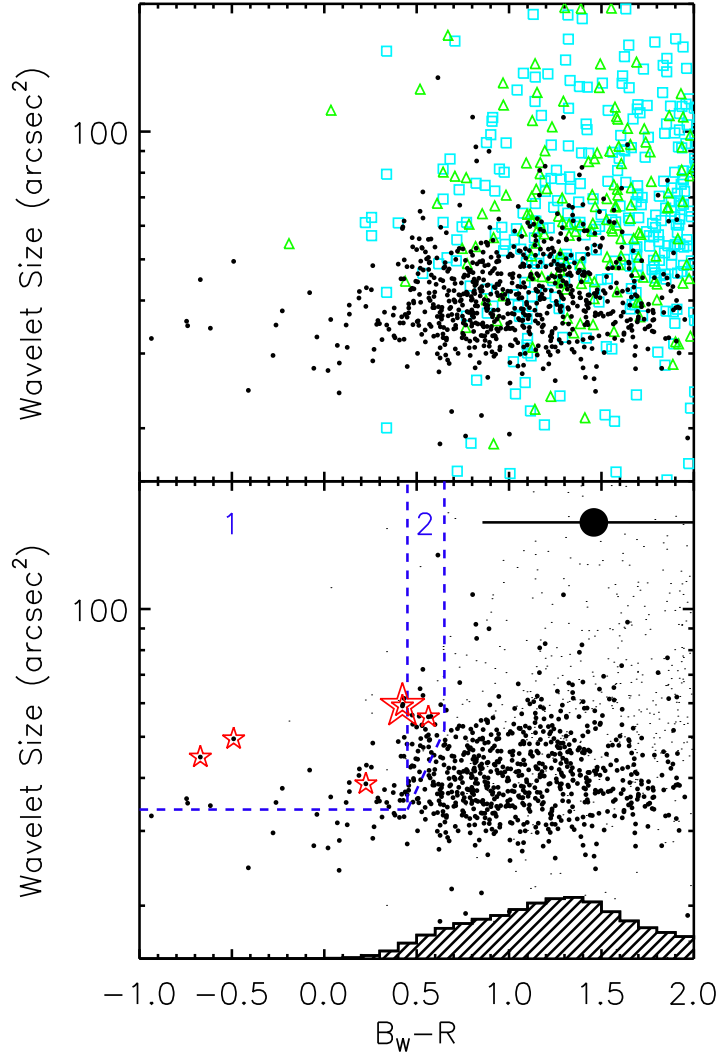}
\caption[Size vs. $B_{W}-R$ of \lya\ Nebula Candidates]
{Wavelet Size vs. $B_{W}-R$ color for \lya\ nebula candidates selected using the morphological search pipeline.  
The wavelet size corresponds to the size of the source in the wavelet power map, but does not indicate the 
true nebular size of the object (Section~\ref{sec:selection}).  
{\bf Top:} \lya\ nebula candidates are shown (black circles) along with artifacts from the residual halos of 
bright galaxies ({\sc galaxy}; cyan squares) and detections of tidal tails and spiral arms ({\sc tidal/arm}; green triangles).
{\bf Bottom:} \lya\ nebula candidates are shown as in the top panel (black circles), with first and second priority selection 
regions indicated (blue dashed lines).  The first (second) priority region contains \numfirst\ (\numsecond) \lya\ nebula candidates.
LABd05 is indicated with a double star \citep{dey05}, and the other spectroscopically-confirmed \lya\ nebulae are 
shown as single stars \citep[][Paper~II]{pres09}.  Artifacts from the upper panel are indicated with small black dots.  
The large filled black circle with an error bar represents the typical color of low surface brightness 
galaxies \citep[LSBs;][]{hab07}, and the 
histogram (plotted on a linear scale) represents the distribution of $B_{W}-R$ colors for field galaxies in NDWFS, 
demonstrating that the colors of our final \lya\ nebula candidates are substantially bluer than typical LSBs and 
field galaxies. 
}
\label{fig:fullsample}
\end{figure}

{\it Step 13:} The final sample was then prioritized based on $B_{W}-R$ color and size, giving priority to 
blue candidates with large sizes.  Figure~\ref{fig:fullsample} shows the size and color distribution of 
the candidate sample.  
We measured $B_{W}-R$ colors in the original images using large apertures (30 pixel, 7.7\arcsec\ diameter, 
i.e., roughly 60~kpc; aperture sky subtraction).
The size of each source, for the purposes of selecting candidates, was taken to 
be the size measured in the wavelet power map ({\sc isoarea\_world} in square degrees), 
as determined by SourceExtractor during the final search ({\it Step 10}).  
While representative, this ``wavelet size'' should not be taken to be the true size of the object.  
The wavelet size typically underestimates the nebular extent in cases where some portion
of that object has been rejected during the pipeline (due to compact sources within or adjacent to
the diffuse emission).  In addition, as we are attempting to detect line emission within the 
broad \bw\ band, the measured \bw\ sizes will underestimate the true nebular size.  
We investigate this in more detail in Paper~II.

\begin{figure}
\center
\includegraphics[angle=0,width=4in]{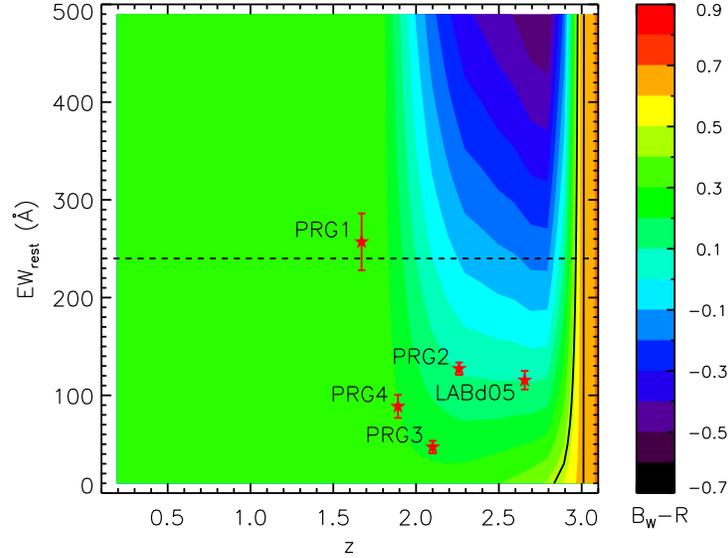}
\caption[Expected $B_{W}-R$ Colors for \lya\ Nebulae.]
{Expected $B_{W}-R$ colors for \lya\ nebulae at redshifts $0.1<z<3.1$, where the colors are
divided into 0.1 mag bins; the $B_{W}-R$ color cuts used in our survey are shown as
black lines at $B_{W}-R<0.45$ and at $B_{W}-R<0.65$.  
\lya\ nebulae are modeled as flat spectrum sources ($f_{\lambda}\propto\lambda^{-2}$) 
with \lya\ emission lines of varying equivalent widths, 
where the continuum at the position of \lya\ is measured using a $20$\AA\ bin redward of the line.  
The \bw\ bandpass contains \lya\ at $1.9\lesssim z \lesssim2.9$, leading to blue colors.  
However, due to the wavelength range of the spectroscopic follow-up observations, \lya\ nebulae can be detected as 
spatially extended sources even at lower redshifts ($z\gtrsim1.6$) if the continuum has the same extended morphology 
as the \lya, even though \lya\ is not contained with the \bw\ band.  
The canonical upper limit for \lya\ equivalent widths arising from stellar
processes is shown \citep[240\AA, dashed line;][]{char93} along with the positions in redshift-EW$_{rest}$ space 
of the confirmed \lya\ nebulae \citep[][Paper~II]{dey05,pres09}.  
The \bw-$R$ colors of actual \lya\ nebulae can and do deviate from the colors predicted using this 
simple model most likely due to, e.g., differing continuum shapes, the composite nature of actual \lya\ 
nebula systems, the presence of embedded sources with different colors.  
}
\label{fig:color}
\end{figure}

The majority of the morphologically-selected candidates are red ($B_{W}-R\gtrsim0.5$), 
but a small subset show blue colors,
which could either result from a strong emission line within the \bw\ band (\lya\ at $1.9<z<2.9$), 
from very blue continuum emission, or a combination.  
Figure \ref{fig:color} shows the expected $B_{W}-R$ color for a model source 
with a given redshift and \lya\ equivalent width generated by adding a Gaussian emission line 
to a flat spectrum source ($f_{\lambda}\propto \lambda^{-2}$).
Unlike narrow-band \lya\ nebula surveys, our survey is not formally equivalent width limited.  
Other than at the high redshift end of the survey ($z\gtrsim2.7$), sources of any equivalent width will
fall within our color selection window as long as they are spatially extended enough and bright enough in 
\bw\ to be selected via the morphological search pipeline.  
However, given the distribution of field galaxy colors we expect that sources with bluer colors are more 
likely to be \lya\ nebulae.  
Given the number of candidates we expected to be able to target during our spectroscopic follow-up 
campaign, we therefore defined our first priority region as $B_{W}-R < 0.45$ and
{\sc isoarea\_world}$>34$ arcsec$^{2}$ and a second priority region as $0.45 < B_{W}-R < 0.65$ and
Log({\sc isoarea\_world})$>0.934\times(B_{W}-R)+1.11$ arcsec$^{2}$ (Figure~\ref{fig:fullsample}).  
Our spectroscopic follow-up observations (using the MMT/Blue Channel Spectrograph with 
a full wavelength range of $\Delta\lambda=3100-8320$\AA) are able to detect line emission from 
\lya\ nebulae down to $z\approx1.6$, leading to a total survey 
redshift range of $1.6<z<2.9$.  The details of our selection function will be treated in Paper~III.

\section{Candidate Sample}
\label{sec:candidate}

Our final candidate sample consists of \numfirst\ first priority and \numsecond\ second priority sources over a search 
area of \searcharea\ square degrees (Table~\ref{tab:select}).  
Of the \numfirst\ first priority sources, \numfirstthree\ have a morphological category of {\sc group} and 
\numfirstfour\ are in the category {\sc diffuse}; for the \numsecond\ second priority sources, \numsecondthree\ are 
in the {\sc group} category and \numsecondfour\ in the {\sc diffuse} category.
The distribution of \lya\ nebulae candidates on the sky is shown in Figure~\ref{fig:footprint}, and postage stamps 
of the \bw, $R$, and $I$-band imaging of all candidates are shown in Appendix~\ref{appendixA}.  
Due to the fact that we included both category {\sc group} and {\sc diffuse} sources in the final catalog, 
the candidates span a range of morphologies from clumpy to diffuse.  
In particular, a number of the very blue sources appear to be a close grouping of compact sources selected as one.
While at first glance these compact groups do not have the diffuse appearance we expect for \lya\ nebulae, 
it could be unwise to remove them because the full morphological range spanned by the \lya\ nebula class is not 
well-known and some \lya\ nebulae are known to include compact sources (e.g., young galaxies; Prescott et al. 2012c, submitted).   
For these reasons we did not apply any further morphological selection criteria to the main sample.  
In addition to our \numcand\ systematically-selected first and second priority samples, we also flagged \numthird\ additional 
candidates with promising morphologies from outside these regions as a test of our color-selection approach.  
Of the third priority sources, which are included for reference in Table~\ref{tab:select} and Appendix~\ref{appendixA}, \numthirdfour\ have a morphological category of {\sc diffuse} and \numthirdthree\ 
is in the category {\sc group}. 

\begin{figure}
\center
\includegraphics[angle=0,width=4in]{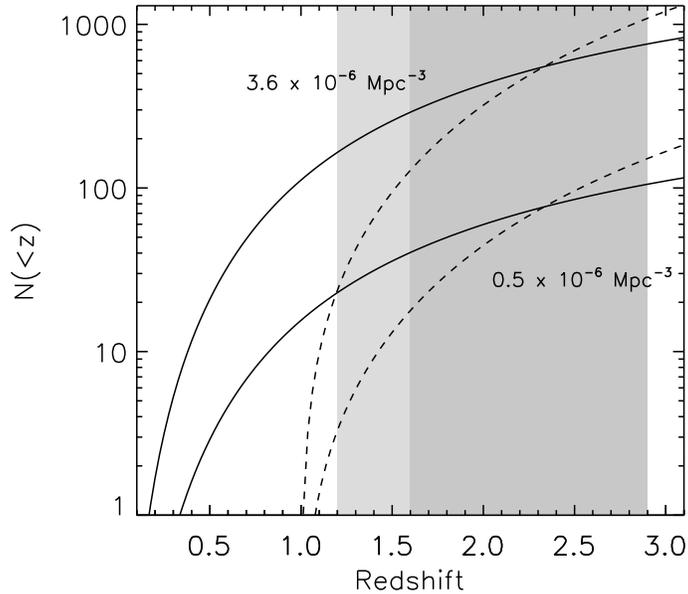}
\caption{Cumulative expected number of \lya\ nebulae assuming a constant (i.e., redshift independent) comoving volume 
density of $0.5\times10^{-6}-3.6\times10^{-6}$ Mpc$^{-3}$ \citep[the range estimated for $z\approx2.3$ by][]{yang09} as 
a function of redshift and a 100\% detection rate (solid lines).  
The dashed lines show the cumulative expected number of \lya\ nebulae under the assumption that the comoving volume 
density increases linearly to higher redshift and matches the measured range at $z=2.3$.  
Our survey is designed to select \lya\ nebulae at redshifts $z\approx1.6-2.9$ (dark grey shading); 
however, \lya\ nebulae at $z\approx1.2-1.6$ (light grey shading) with diffuse blue continuum emission that are 
selected by our morphological survey will appear as continuum-only sources in follow-up spectroscopy. 
}
\label{fig:expblobs}
\end{figure}

Based on previous narrow-band \lya\ nebulae surveys, we can predict the number of \lya\ nebulae we would expect to find within our full survey volume.  
Figure~\ref{fig:expblobs} shows the cumulative number of \lya\ nebulae expected assuming a constant comoving volume density as a function of redshift of $0.5\times10^{-6}-3.6\times10^{-6}$ Mpc$^{-3}$ 
\citep[the range spanned by the estimates of][]{yang09} and a 100\% detection success 
rate.  This non-evolving model would predict approximately 60-400 \lya\ nebulae over 
the redshift range of our survey ($1.6\lesssim z \lesssim2.9$).  
As the comoving volume density of \lya\ nebulae is unlikely to remain constant over 
cosmic time \citep{keel09}, for comparison we also plot the cumulative number of \lya\ nebulae expected 
within our survey area assuming that the comoving volume density increases linearly as 
a function of redshift (dashed lines).  
In reality the selection function, completeness, and efficiency of our survey 
are different from that of narrow-band surveys such as \citet{yang09} and the 
evolution of the \lya\ nebula comoving volume density is unknown; we address 
these issues in more detail in Paper~III.

\subsection{Potential Contaminants}
\label{sec:potcontam}
Due to the wide bandpass of the \bw\ filter, there are several populations 
that may contaminate our \lya\ nebula candidate sample. 
The morphological selection method will tend to pick out any source of low surface brightness 
extended emission that cannot be easily identified, e.g., a low surface brightness galaxy (LSB) 
or low surface brightness nebulae within the Galaxy.  
In principle, lower redshift sources with \oii$\lambda\lambda$3727,3729 emission in the selection 
band (\bw) are a contaminant population, much as they are in \lya-emitting galaxy surveys.  
In addition, galaxies and \lya\ nebulae in the redshift desert (i.e., $1.2<z<1.6$) 
may be selected as candidates in our survey if they exhibit extended blue continuum emission or 
a close grouping of compact sources.  In these cases, \lya\ would be positioned 
blueward and \oii\ redward of the wavelength range of our follow-up spectroscopy, leading us to detect 
faint, spatially extended, blue continuum emission with no emission lines to provide confirmation of the redshift.  
 
While this range of different populations could contaminate our sample in principle, 
not all of them will contribute substantially in practice.  We do not expect to find many LSBs because 
the $B_{W}-R$ colors of \lya\ nebulae are likely to be bluer than typical LSB colors 
\citep[i.e., $B_{W}-R=1.49\pm0.60$;][]{hab07}; any blue LSBs would be expected to show 
optical emission lines and therefore easily identified in follow-up spectroscopy.  
If compact groupings of \oii-emitters or extended \oii\ nebulae were to be selected 
via our morphological search, the position of \oii\ in the \bw, \oiii\ in the $R$ band, and \halpha\ 
in the $R$ or $I$ band would give the object redder colors, 
which would not be included in the final sample.  
On the other hand, a primary contaminant population will be galaxies or \lya\ nebulae in 
the redshift range $1.2\lesssim z\lesssim 1.6$ that appear in optical spectroscopy as spatially extended 
continuum-only sources.  From Figure~\ref{fig:expblobs}, we would expect of order 20-180 
such sources, or roughly 25\% of the predicted 
\lya\ nebulae sample over the entire redshift range ($1.2\lesssim z \lesssim 2.9$), 
assuming no evolution in the intrinsic number density.  
With line emission outside the optical window, however, follow-up observations with HST will 
be necessary to confirm whether or not any continuum-only sources are in fact lower redshift \lya\ nebulae.

\section{Discussion}
\label{sec:discussion}

\subsection{A Successful Broad-band Ly$\alpha$ Nebula Survey}
\label{sec:bbsurvey}

\begin{figure}
\center
\includegraphics[angle=0,width=4in]{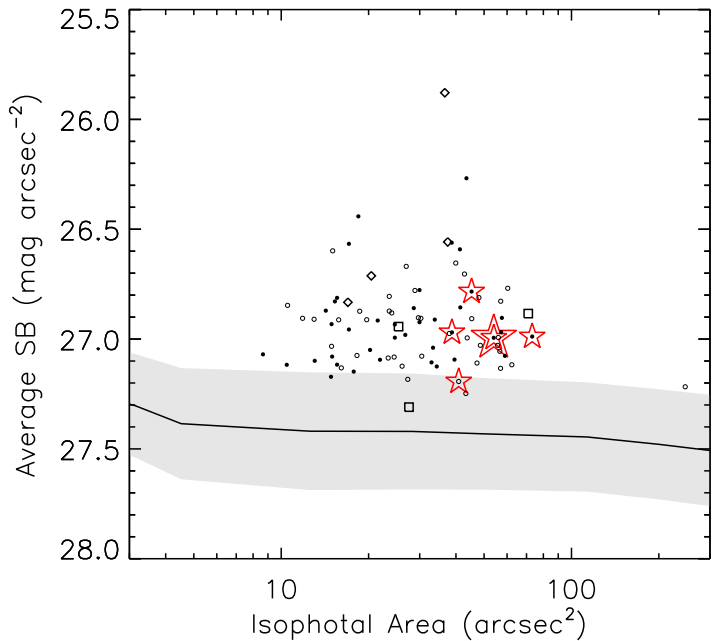}
\caption{
Average surface brightness versus isophotal area (above the median surface brightness limit of the NDWFS survey; \sblimonesig\ 
mag arcsec$^{2}$, 1$\sigma$, 1.1\arcsec\ diameter aperture) for both first (filled circles) and second (open circles) 
priority candidates.  Confirmed \lya\ nebulae from this survey are indicated with stars (Paper~II), 
including LABd05 \citep[double star;][]{dey05}.  
Spectroscopically confirmed \lya\ sources in the Bo\"otes field from other surveys that were not 
selected by our broad-band search (Table~\ref{tab:otherblob}), primarily due to the presence of bright 
central sources, are shown as diamonds \citep{yang09} and squares (Prescott et al. 2012e, in prep.).  
The median 5$\sigma$ surface brightness limit of the survey is shown as the black line, with the standard deviation of 
all 27 pointings shown as a shaded band. 
}
\label{fig:SBsize}
\end{figure}

Our systematic broad-band \lya\ nebula search was designed to find luminous \lya\ nebulae 
similar to LABd05.  
Subsequent spectroscopic follow-up of a subset of our morphological candidate sample 
showed that, in addition to recovering the previously-discovered large \lya\ nebula at
$z\approx2.66$ \citep[LABd05;][]{dey05}, this novel approach was able to successfully 
locate 4 new large \lya\ nebulae over the NDWFS Bo\"otes field at $z\approx1.67$, $z\approx2.27$, 
$z\approx2.14$, and $z\approx1.88$ \citep[PRG1, PRG2, PRG3, PRG4;][]{presphd}.  
Of particular note, our broad-band search identified the first spatially extended \lya+\heii\ nebula 
\citep[PRG1;][]{pres09}; located at $z\approx1.67$, this system is the lowest redshift \lya\ nebula known, 
and the strong \heii\ emission suggests that it may contain low metallicity gas.  
The 5 confirmed \lya\ nebulae are shown for reference in Figures~\ref{fig:fullsample}-\ref{fig:color}, 
\ref{fig:footprint}, and \ref{fig:SBsize}.  
A full discussion of the spectroscopic follow-up campaign is presented in Paper~II, and 
the selection function and implied space density for this survey are discussed in Paper~III.
 
\subsection{Comparisons to Other Surveys}
\label{sec:comparison}

While broad-band imaging has been used in recent years to isolate samples of \lya-emitting and \lya-absorbing 
galaxies based on colors \citep{cooke09} and to find strong line-emitting nebulae at low redshifts 
through the Galaxy Zoo project \citep{lintott09,keel11}, our search is the first time broad-band data has been 
used as the basis for a semi-automated systematic morphological survey for high redshift \lya\ nebulae.  
Our \lya\ nebula survey is also fundamentally different and complementary to 
standard narrow-band surveys for \lya\ nebulae (Table~\ref{tab:othersearches}).  
The pilot survey of the NDWFS Bo\"otes field has shown that it is possible to find 
large \lya\ nebulae using deep broad-band imaging data.  
There are many advantages to this unusual approach.
A broad-band search is able to efficiently cover enormous comoving volumes that would be 
observationally expensive when using a narrow-band filter.
A narrow-band survey using a 100\AA\ filter at $z\approx2.3$ would need to search
over 100 square degrees to reach the same comoving volume covered in our broad-band survey of 
the NDWFS Bo\"otes field ($\approx10^{8}$ h$_{70}^{3}$ Mpc$^{-3}$).
In addition, our technique builds on existing datasets, which means that our search algorithm can be easily 
modified to find \lya\ nebulae within any deep, broad-band, wide-field imaging survey, leading to 
substantially lower observational overhead.  
Finally, while the early, successful \lya\ surveys often targeted known overdensities and were
limited to smaller areas \citep{stei00,mat04,francis96,pal04},
introducing a bias to the number density estimates, our \searcharea\ square degree survey, by contrast, 
is unbiased in terms of environment, spanning a $\Delta z\approx1.3$, a comoving transverse distance of $\approx$290 h$_{70}^{-1}$ Mpc 
(at the redshift midpoint, $z\approx2.25$), and a comoving line-of-sight distance of $\approx$1700 h$_{70}^{-1}$ Mpc.

At the same time, a drawback of using broad-band data to select line-emitting sources is the necessity 
of restricting our search to the largest, most luminous \lya\ nebulae and of removing compact sources 
in favor of diffuse emission, in order to reduce the number of contaminants.  
Thus, while our search successfully recovered a sample of \lya\ nebulae, a handful of other 
(fainter or more compact) \lya\ nebulae that were identified by traditional narrow-band searches 
in Bo\"otes were not recovered by our broad-band search (Table~\ref{tab:otherblob}).  

In Figure~\ref{fig:SBsize}, 
we show the \bw\ isophotal area and mean surface brightness estimates for the candidate sample, as 
derived using SourceExtractor ({\sc detect\_thresh}$=\sblimonesig$ mag arcsec$^{-2}$, 
{\sc detect\_minarea}$=5$) on the original \bw\ imaging, for the \lya\ nebulae confirmed by our survey 
(Paper~II), and for other known \lya\ nebula sources in the field.  
The sources our survey did not recover fall into two categories: either
the \lya\ emission surrounds a much brighter compact continuum source (``core-halo'' morphology) 
or the diffuse \bw\ emission is simply too faint to be selected by our search algorithm.  
In the first category are the four \lya\ nebulae found by \citet{yang09} using a traditional narrow-band survey 
of NDWFS Bo\"otes; all four consist of bright continuum sources (AGN in two cases) surrounded by \lya\ halos that are 
undetected in the broad-band \bw\ imaging.  The bright compact central sources caused these 
objects to be rejected and replaced with sky noise during Steps 2-3 or Steps 6-7 of the search pipeline.
Intermediate-band imaging with the Subaru \citep{pres08} and Mayall Telescopes 
revealed three additional \lya\ nebulae that were not detected using the 
broad-band search algorithm.  One case (P1) was too faint to be selected by our search algorithm, and 
two other cases (P2 \& P3) showed core-halo morphologies with confirmed \lya\ emission 
(at $z\approx2.6$ and $z\approx1.9$) in the midst of several bright continuum sources.  
With the compact sources removed by the search pipeline, 
the wings of the \lya\ emission were still detected in 
the \bw\ imaging in each case, but at 
weaker wavelet power limits than probed by the current survey.  

As our search was designed to select large, luminous, and diffuse \lya\ nebulae, 
it is not surprising that more compact or fainter \lya\ nebulae were not selected.  
In future work, it will be important to understand this range of \lya\ 
nebula morphologies and whether they reflect fundamentally different 
sources or simply different evolutionary phases of the same underlying 
phenomenon.


\section{Conclusions}
\label{sec:conclusions}

We have designed an innovative systematic search for large \lya\ nebulae using 
deep broad-band data.  Our technique is designed to find the largest and brightest \lya\ 
nebulae and is able to probe enormous comoving volumes ($\approx10^{8}$ h$^{-3}_{70}$ Mpc$^{3}$) 
using existing deep broad-band datasets.  In this paper, we presented details on the survey design and our sample of 
\lya\ nebula candidates selected from the NDWFS Bo\"otes Field.  Paper~II of this series presents 
the spectroscopic follow-up and the confirmed \lya\ nebula sample, and 
Paper~III will address the details of the survey selection function and the implications 
for the space density of \lya\ nebulae.  
In the future, we will build on the search algorithm developed in this work in order to find \lya\ nebulae 
in other deep and wide imaging fields and increase the sample of known \lya\ nebulae available for 
detailed follow-up study.

\acknowledgments
The authors are grateful to Alexey Vikhlinin for the use of the code {\it wvdecomp}.  
This research draws upon data from the NDWFS as distributed by the NOAO Science Archive. 
NOAO is operated by the Association of Universities for Research in Astronomy (AURA), Inc. under 
a cooperative agreement with the National Science Foundation.  
M. P. was supported by a NSF Graduate Research Fellowship and a TABASGO Prize Postdoctoral Fellowship.  
A. D. and B. T. J.'s research is supported by NOAO, which is operated by AURA under a cooperative 
agreement with the National Science Foundation.  


\renewcommand{\thefootnote}{\alph{footnote}}
\begin{longtable}{ccccccccc}
\tabletypesize{\scriptsize}
\tablecaption{Ly$\alpha$ Nebula Candidates}
\tablewidth{0pt}
\tablehead{
 & \colhead{Candidate Name} & \colhead{Right Ascension} & \colhead{Declination} & \colhead{Wavelet Size} & \colhead{Morphological} & \colhead{$B_{W}$\tablenotemark{b}} & \colhead{$B_{W}-R$\tablenotemark{b}} & \colhead{Priority\tablenotemark{c}} \\
 & & (J2000 hrs) & (J2000 deg) & (arcsec$^{2})$ & Flag\tablenotemark{a} & (AB mag) & (AB mag) & \\
 }
     1) &                   NDWFS J143004.7+353509 & 14:30:04.687 & 35:35:09.06 & 134.0 &        {\sc diffuse} & 23.79$\pm$0.05 &  0.61 & 2\tablenotemark{d} \\ 
     2) &                   NDWFS J143006.9+353437 & 14:30:06.864 & 35:34:36.73 & 108.2 &        {\sc diffuse} & 23.05$\pm$0.03 &  0.80 & 3\tablenotemark{d} \\ 
     3) &                   NDWFS J142846.2+330819 & 14:28:46.228 & 33:08:19.42 &  89.9 &        {\sc diffuse} & 23.00$\pm$0.03 &  0.89 & 3 \\ 
     4) &                   NDWFS J143436.4+330406 & 14:34:36.448 & 33:04:05.62 &  72.1 &          {\sc group} & 22.90$\pm$0.02 &  0.53 & 2 \\ 
     5) &                   NDWFS J142855.0+353022 & 14:28:54.998 & 35:30:21.70 &  64.9 &        {\sc diffuse} & 23.26$\pm$0.03 &  0.52 & 2 \\ 
     6) &                   NDWFS J143153.1+351436 & 14:31:53.066 & 35:14:36.45 &  62.6 &          {\sc group} & 23.31$\pm$0.04 &  0.53 & 2 \\ 
     7) &                   NDWFS J143055.2+352855 & 14:30:55.192 & 35:28:54.76 &  61.6 &          {\sc group} & 23.30$\pm$0.04 &  0.43 & 1 \\ 
     8) &                   NDWFS J143057.7+354547 & 14:30:57.662 & 35:45:47.26 &  59.6 &          {\sc group} & 23.09$\pm$0.03 &  0.43 & 1 \\ 
     9) &                   NDWFS J143842.0+335325 & 14:38:41.956 & 33:53:24.79 &  59.5 &          {\sc group} & 23.14$\pm$0.04 &  0.63 & 2 \\ 
    10) &                   NDWFS J143410.9+331731 & 14:34:10.948 & 33:17:30.80 &  59.1 &        {\sc diffuse} & 23.18$\pm$0.03 &  0.42 & 1\tablenotemark{e} \\ 
    11) &                   NDWFS J142729.9+341125 & 14:27:29.894 & 34:11:25.22 &  56.2 &          {\sc group} & 23.48$\pm$0.05 &  0.48 & 2 \\ 
    12) &                   NDWFS J143551.3+333421 & 14:35:51.278 & 33:34:21.03 &  55.9 &          {\sc group} & 23.38$\pm$0.05 &  0.51 & 2 \\ 
    13) &                   NDWFS J142909.3+342542 & 14:29:09.278 & 34:25:42.38 &  55.8 &          {\sc group} & 23.41$\pm$0.04 &  0.58 & 2 \\ 
    14) &                   NDWFS J143512.3+351109 & 14:35:12.336 & 35:11:08.62 &  55.7 &        {\sc diffuse} & 23.53$\pm$0.04 &  0.56 & 2\tablenotemark{f} \\ 
    15) &                   NDWFS J142906.3+341130 & 14:29:06.271 & 34:11:30.26 &  53.5 &          {\sc group} & 23.47$\pm$0.04 &  0.48 & 2 \\ 
    16) &                   NDWFS J142856.0+331937 & 14:28:56.013 & 33:19:36.51 &  53.5 &          {\sc group} & 23.38$\pm$0.04 &  0.60 & 2 \\ 
    17) &                   NDWFS J143511.0+335543 & 14:35:11.030 & 33:55:43.10 &  53.3 &          {\sc group} & 23.12$\pm$0.03 &  0.53 & 2 \\ 
    18) &                   NDWFS J143222.8+324943 & 14:32:22.768 & 32:49:42.67 &  53.1 &        {\sc diffuse} & 23.35$\pm$0.04 &  0.53 & 2 \\ 
    19) &                   NDWFS J143302.8+354729 & 14:33:02.844 & 35:47:29.43 &  52.7 &          {\sc group} & 23.42$\pm$0.04 &  0.50 & 2 \\ 
    20) &                   NDWFS J143641.4+341510 & 14:36:41.373 & 34:15:10.26 &  52.1 &          {\sc group} & 23.50$\pm$0.04 &  0.55 & 2 \\ 
    21) &                   NDWFS J143230.2+334105 & 14:32:30.244 & 33:41:05.46 &  50.5 &          {\sc group} & 23.25$\pm$0.03 &  0.56 & 2 \\ 
    22) &                   NDWFS J142819.8+344657 & 14:28:19.840 & 34:46:57.14 &  50.4 &          {\sc group} & 23.16$\pm$0.03 &  0.61 & 2 \\ 
    23) &                   NDWFS J143115.1+354148 & 14:31:15.055 & 35:41:47.76 &  50.1 &          {\sc group} & 23.54$\pm$0.05 &  0.55 & 2 \\ 
    24) &                   NDWFS J142614.7+344434 & 14:26:14.714 & 34:44:34.22 &  49.9 &          {\sc group} & 23.51$\pm$0.05 &  0.43 & 1 \\ 
    25) &                   NDWFS J142929.2+354711 & 14:29:29.198 & 35:47:10.78 &  49.5 &          {\sc group} & 23.17$\pm$0.03 &  0.52 & 2 \\ 
    26) &                   NDWFS J142622.9+351422 & 14:26:22.905 & 35:14:22.02 &  49.5 &        {\sc diffuse} & 23.20$\pm$0.04 & -0.49 & 1\tablenotemark{g} \\ 
    27) &                   NDWFS J143055.7+340502 & 14:30:55.713 & 34:05:01.71 &  49.3 &        {\sc diffuse} & 23.04$\pm$0.03 &  0.57 & 2 \\ 
    28) &                   NDWFS J143236.9+351800 & 14:32:36.883 & 35:18:00.32 &  49.1 &          {\sc group} & 23.42$\pm$0.06 &  0.57 & 2 \\ 
    29) &                   NDWFS J142526.3+335112 & 14:25:26.332 & 33:51:12.16 &  48.7 &          {\sc group} & 23.35$\pm$0.04 &  0.40 & 1 \\ 
    30) &                   NDWFS J143407.5+334141 & 14:34:07.476 & 33:41:40.56 &  48.3 &          {\sc group} & 23.43$\pm$0.05 &  0.26 & 1 \\ 
    31) &                   NDWFS J142547.1+334454 & 14:25:47.126 & 33:44:54.13 &  48.1 &        {\sc diffuse} & 22.98$\pm$0.03 &  0.45 & 1 \\ 
    32) &                   NDWFS J143459.7+333749 & 14:34:59.702 & 33:37:48.93 &  48.1 &          {\sc group} & 23.68$\pm$0.06 &  0.48 & 2 \\ 
    33) &                   NDWFS J142714.8+343155 & 14:27:14.791 & 34:31:54.55 &  47.7 &          {\sc group} & 23.38$\pm$0.04 &  0.52 & 2 \\ 
    34) &                   NDWFS J143128.2+352658 & 14:31:28.245 & 35:26:57.91 &  47.3 &        {\sc diffuse} & 23.06$\pm$0.03 &  0.56 & 2 \\ 
    35) &                   NDWFS J142906.4+352432 & 14:29:06.398 & 35:24:32.22 &  47.2 &          {\sc group} & 23.50$\pm$0.05 &  0.54 & 2 \\ 
    36) &                   NDWFS J143422.3+351534 & 14:34:22.348 & 35:15:34.27 &  46.3 &          {\sc group} & 23.50$\pm$0.04 &  0.51 & 2 \\ 
    37) &                   NDWFS J143310.3+335057 & 14:33:10.346 & 33:50:56.94 &  45.9 &          {\sc group} & 23.82$\pm$0.07 &  0.58 & 2 \\ 
    38) &                   NDWFS J143523.9+354934 & 14:35:23.856 & 35:49:34.39 &  45.0 &          {\sc group} & 23.93$\pm$0.07 &  0.42 & 1 \\ 
    39) &                   NDWFS J143053.5+352007 & 14:30:53.488 & 35:20:06.68 &  45.0 &          {\sc group} & 23.02$\pm$0.04 &  0.43 & 1 \\ 
    40) &                   NDWFS J142653.2+343855 & 14:26:53.172 & 34:38:55.39 &  44.8 &          {\sc group} & 23.53$\pm$0.06 & -0.67 & 1\tablenotemark{i} \\ 
    41) &                   NDWFS J142758.6+354428 & 14:27:58.617 & 35:44:28.42 &  44.4 &          {\sc group} & 23.42$\pm$0.03 &  0.50 & 2 \\ 
    42) &                   NDWFS J142634.1+334954 & 14:26:34.106 & 33:49:53.61 &  44.3 &          {\sc group} & 23.54$\pm$0.04 &  0.51 & 2 \\ 
    43) &                   NDWFS J143435.9+350710 & 14:34:35.932 & 35:07:09.91 &  44.2 &          {\sc group} & 23.72$\pm$0.05 &  0.57 & 2 \\ 
    44) &                   NDWFS J142927.8+345906 & 14:29:27.837 & 34:59:06.14 &  43.1 &          {\sc group} & 23.18$\pm$0.03 &  1.18 & 3 \\ 
    45) &                   NDWFS J142745.2+352949 & 14:27:45.194 & 35:29:48.98 &  42.9 &          {\sc group} & 23.48$\pm$0.04 &  0.23 & 1 \\ 
    46) &                   NDWFS J142826.0+352116 & 14:28:25.999 & 35:21:15.62 &  42.7 &          {\sc group} & 23.58$\pm$0.04 &  0.48 & 2 \\ 
    47) &                   NDWFS J142535.2+324934 & 14:25:35.205 & 32:49:34.17 &  42.6 &          {\sc group} & 23.54$\pm$0.05 &  0.49 & 2 \\ 
    48) &                   NDWFS J142960.0+353927 & 14:29:59.978 & 35:39:26.89 &  42.4 &          {\sc group} & 23.36$\pm$0.03 &  0.47 & 2 \\ 
    49) &                   NDWFS J142821.3+352001 & 14:28:21.259 & 35:20:00.92 &  42.4 &          {\sc group} & 23.42$\pm$0.04 &  0.25 & 1 \\ 
    50) &                   NDWFS J143401.0+351229 & 14:34:01.034 & 35:12:29.12 &  42.0 &          {\sc group} & 23.91$\pm$0.06 &  0.19 & 1 \\ 
    51) &                   NDWFS J143348.9+354157 & 14:33:48.885 & 35:41:56.68 &  41.7 &          {\sc group} & 23.74$\pm$0.05 & -0.08 & 1 \\ 
    52) &                   NDWFS J143706.6+335653 & 14:37:06.588 & 33:56:52.65 &  41.3 &        {\sc diffuse} & 23.22$\pm$0.03 &  0.52 & 2 \\ 
    53) &                   NDWFS J142446.0+354714 & 14:24:45.998 & 35:47:14.06 &  40.5 &          {\sc group} & 23.68$\pm$0.05 &  0.19 & 1 \\ 
    54) &                   NDWFS J142522.3+325424 & 14:25:22.339 & 32:54:23.65 &  39.9 &          {\sc group} & 23.40$\pm$0.04 &  0.48 & 2 \\ 
    55) &                   NDWFS J143833.2+340359 & 14:38:33.249 & 34:03:59.32 &  39.6 &          {\sc group} & 23.74$\pm$0.06 &  0.52 & 2 \\ 
    56) &                   NDWFS J143141.3+352110 & 14:31:41.253 & 35:21:09.86 &  39.5 &        {\sc diffuse} & 23.17$\pm$0.05 &  0.48 & 2 \\ 
    57) &                   NDWFS J143416.4+333058 & 14:34:16.420 & 33:30:58.10 &  39.3 &          {\sc group} & 23.26$\pm$0.05 &  0.46 & 2 \\ 
    58) &                   NDWFS J142516.6+324335 & 14:25:16.629 & 32:43:35.47 &  39.0 &          {\sc group} & 23.13$\pm$0.04 &  0.14 & 1 \\ 
    59) &                   NDWFS J143412.7+332939 & 14:34:12.722 & 33:29:39.19 &  38.7 &        {\sc diffuse} & 23.40$\pm$0.07 &  0.22 & 1\tablenotemark{h} \\ 
    60) &                   NDWFS J142441.9+332532 & 14:24:41.940 & 33:25:31.94 &  38.4 &          {\sc group} & 23.52$\pm$0.04 &  0.38 & 1 \\ 
    61) &                   NDWFS J143438.7+350420 & 14:34:38.700 & 35:04:19.84 &  38.4 &        {\sc diffuse} & 23.72$\pm$0.05 &  0.46 & 2 \\ 
    62) &                   NDWFS J142832.7+354105 & 14:28:32.733 & 35:41:04.56 &  37.8 &          {\sc group} & 23.72$\pm$0.05 & -0.23 & 1 \\ 
    63) &                   NDWFS J143335.8+344503 & 14:33:35.764 & 34:45:02.84 &  37.5 &          {\sc group} & 23.92$\pm$0.06 &  0.33 & 1 \\ 
    64) &                   NDWFS J143203.8+351855 & 14:32:03.760 & 35:18:54.86 &  37.4 &          {\sc group} & 23.46$\pm$0.05 &  0.46 & 2 \\ 
    65) &                   NDWFS J143207.2+343101 & 14:32:07.224 & 34:31:01.34 &  37.3 &        {\sc diffuse} & 23.30$\pm$0.03 &  0.81 & 3 \\ 
    66) &                   NDWFS J142539.9+344959 & 14:25:39.859 & 34:49:59.19 &  37.1 &          {\sc group} & 23.57$\pm$0.04 &  0.14 & 1 \\ 
    67) &                   NDWFS J142819.7+325449 & 14:28:19.735 & 32:54:48.70 &  37.1 &          {\sc group} & 23.44$\pm$0.03 &  0.44 & 1 \\ 
    68) &                   NDWFS J142443.9+344834 & 14:24:43.869 & 34:48:34.45 &  36.9 &          {\sc group} & 23.64$\pm$0.05 &  0.37 & 1 \\ 
    69) &                   NDWFS J143202.6+351904 & 14:32:02.568 & 35:19:04.22 &  36.6 &        {\sc diffuse} & 22.99$\pm$0.04 &  0.42 & 1 \\ 
    70) &                   NDWFS J142753.8+341204 & 14:27:53.762 & 34:12:04.10 &  36.5 &        {\sc diffuse} & 23.29$\pm$0.04 &  0.45 & 1 \\ 
    71) &                   NDWFS J142600.8+350252 & 14:26:00.842 & 35:02:52.36 &  36.1 &        {\sc diffuse} & 23.87$\pm$0.06 &  0.63 & 3 \\ 
    72) &                   NDWFS J142643.9+340937 & 14:26:43.850 & 34:09:36.82 &  36.1 &          {\sc group} & 23.40$\pm$0.03 &  0.03 & 1 \\ 
    73) &                   NDWFS J142722.4+345225 & 14:27:22.408 & 34:52:24.74 &  35.9 &          {\sc group} & 23.71$\pm$0.05 &  0.29 & 1 \\ 
    74) &                   NDWFS J142620.0+340427 & 14:26:19.982 & 34:04:27.01 &  35.7 &          {\sc group} & 23.51$\pm$0.04 & -0.75 & 1 \\ 
    75) &                   NDWFS J142707.8+344749 & 14:27:07.840 & 34:47:48.84 &  35.4 &        {\sc diffuse} & 23.51$\pm$0.05 &  0.43 & 1 \\ 
    76) &                   NDWFS J142710.4+324842 & 14:27:10.447 & 32:48:41.76 &  35.4 &          {\sc group} & 23.09$\pm$0.04 &  0.41 & 1 \\ 
    77) &                   NDWFS J143232.5+351534 & 14:32:32.484 & 35:15:34.23 &  35.3 &          {\sc group} & 23.45$\pm$0.07 &  0.30 & 1 \\ 
    78) &                   NDWFS J143651.3+342107 & 14:36:51.312 & 34:21:07.38 &  35.1 &          {\sc group} & 23.71$\pm$0.05 &  0.28 & 1 \\ 
    79) &                   NDWFS J142746.3+344544 & 14:27:46.348 & 34:45:44.02 &  35.0 &          {\sc group} & 23.39$\pm$0.04 &  0.34 & 1 \\ 
    80) &                   NDWFS J142548.3+322957 & 14:25:48.283 & 32:29:56.58 &  35.0 &          {\sc group} & 23.83$\pm$0.06 &  0.12 & 1 \\ 
    81) &                   NDWFS J142735.5+342332 & 14:27:35.479 & 34:23:32.38 &  35.0 &          {\sc group} & 23.55$\pm$0.04 & -0.26 & 1 \\ 
    82) &                   NDWFS J142449.8+324743 & 14:24:49.761 & 32:47:42.61 &  34.9 &          {\sc group} & 24.09$\pm$0.09 & -0.74 & 1 \\ 
    83) &                   NDWFS J142732.5+341213 & 14:27:32.520 & 34:12:13.39 &  34.4 &          {\sc group} & 23.62$\pm$0.05 & -0.62 & 1 \\ 
    84) &                   NDWFS J142802.8+350933 & 14:28:02.803 & 35:09:33.30 &  34.2 &          {\sc group} & 23.80$\pm$0.06 &  0.35 & 1 \\ 
    85) &                   NDWFS J142533.0+343912 & 14:25:32.966 & 34:39:11.95 &  28.8 &        {\sc diffuse} & 23.67$\pm$0.06 &  0.26 & 3 \\ 
\footnotetext[1]{Morphological flag discussed in Section~\ref{sec:finalcat}: {\sc group} - tight grouping of compact sources, {\sc diffuse} - spatially-extended diffuse emission.}
\footnotetext[2]{Magnitudes and colors were measured using large 30~pix (7.7\arcsec) diameter apertures.}
\footnotetext[3]{Priority assignments are discussed in Section~\ref{sec:selection}.}
\footnotetext[4]{The two largest candidates are in fact part of a single diffuse, asymmetric, and mostly linear structure.  The unusual morphology and large size suggest that this is more likely some sort of Galactic nebula rather than an extragalactic source.  The optical spectrum (Paper~II) does not show any obvious emission features, suggesting that this might be a reflection nebula.}
\footnotetext[5]{Our search recovered LABd05, the \lya\ nebula found previously by \citet{dey05}.  The listed coordinates are those returned by the search algorithm.}
\footnotetext[6]{Our search discovered PRG1, a \lya\ nebula at $z\approx1.67$ \citep[][Paper~II]{pres09}.}
\footnotetext[7]{Our search discovered PRG2, a \lya\ nebula at $z\approx2.27$ \citep[][Paper~II]{presphd}.}
\footnotetext[8]{Our search discovered PRG3, a \lya\ nebula at $z\approx2.14$ \citep[][Paper~II]{presphd}.}
\footnotetext[9]{Our search discovered PRG4, a \lya\ nebula at $z\approx1.88$ \citep[][Paper~II]{presphd}.}
\label{tab:select}
\end{longtable}

\renewcommand{\thefootnote}{\arabic{footnote}}
\clearpage

\begin{deluxetable}{ccccccc}
\tabletypesize{\scriptsize}
\tablecaption{Wide-Area \lya\ Nebula Surveys}
\tablewidth{0pt}
\tablehead{
\colhead{Survey} & \colhead{Field\tablenotemark{a}} & \colhead{Redshift} & \colhead{$\Delta$z} & \colhead{Area} & \colhead{Comoving Volume} \\
 &  &  &  & (deg$^{2}$) &  ($10^{6}$ h$_{70}^{-3}$ Mpc$^{3}$) \\
 }
\startdata
     \citealt{mat04} &                SSA22 &  3.09 &  0.06 &  0.20 &  0.14 \\
     \citealt{pal04} &           J2143-4423 &  2.38 &  0.04 &  0.56 &  0.30 \\
     \citealt{sai06} &                 SXDS &  3.34 &  0.20 &  0.23 &  0.53 \\
                     &                 SXDS &  3.72 &  0.22 &  0.23 &  0.56 \\
                     &                 SXDS &  3.93 &  0.24 &  0.23 &  0.60 \\
                     &                 SXDS &  4.12 &  0.24 &  0.23 &  0.59 \\
                     &                 SXDS &  4.35 &  0.26 &  0.23 &  0.62 \\
                     &                 SXDS &  4.58 &  0.28 &  0.23 &  0.65 \\
                     &                 SXDS &  4.82 &  0.26 &  0.23 &  0.59 \\
     \citealt{smi07} &    XMM-LSS, LH, SFLS &  2.85 &  0.08 & $\approx$15\tablenotemark{b} & $\approx$14.7 \\
                     &    XMM-LSS, LH, SFLS &  3.00 &  0.14 & $\approx$15\tablenotemark{b} & $\approx$24.8 \\
                     &    XMM-LSS, LH, SFLS &  3.12 &  0.08 & $\approx$15\tablenotemark{b} & $\approx$14.5 \\
     \citealt{mat09} &         B3J2330+3927 &  3.09 &  0.06 &  0.25 &  0.18 \\
     \citealt{ouc09} &                 SXDS &  6.56 &  0.10 &  1.00 &  0.82 \\
    \citealt{yang09} &       NDWFS Bo\"otes &  2.30 &  0.04 &  4.82 &  2.14 \\
    \citealt{yang10} &                CDF-N &  2.30 &  0.04 &  0.28 &  0.12 \\
                     &                CDF-S &  2.30 &  0.04 &  0.24 &  0.11 \\
                     &              COSMOS1 &  2.30 &  0.04 &  0.36 &  0.16 \\
                     &              COSMOS2 &  2.30 &  0.04 &  0.36 &  0.16 \\
     \citealt{mat11} &                SSA22 &  3.09 &  0.06 &  1.06 &  0.74 \\
                     &                 SXDS &  3.09 &  0.06 &  0.60 &  0.42 \\
                     &              GOODS-N &  3.09 &  0.06 &  0.24 &  0.17 \\
                     &                  SDF &  3.09 &  0.06 &  0.22 &  0.16 \\
\hline
  & & & &  & \\
This work & NDWFS Bo\"otes & 1.6-2.9 & 1.3 & 8.5 & 130 \\
\enddata
\tablenotetext{a}{Abbreviated field names correspond to the following: COSMOS1 \& COSMOS2 - two pointings within the Cosmic Evolution Survey (COSMOS) field, CDF-N - Chandra Deep Field North, CDF-S - Chandra Deep Field South, B3J2330+3927 - a high redshift radio galaxy field, GOODS-N - Great Observatories Origins Deep Survey North field, J2143-4423 - a galaxy cluster field, LH - Lockman Hole field, NDWFS Bo\"otes - NOAO Deep Wide-Field Survey Bo\"otes field, SDF - Subaru Deep Field, SFLS - Spitzer First Look Survey field, SSA22 - Small Selected Area 22 field originally from \citet{lilly91}, SXDS - Subaru/XMM-Newton Deep Survey, XMM-LSS - XMM-Newton Large Scale Structure Survey field.}
\tablenotetext{b}{Exact survey area was not given in \citet{smi07}.}
\label{tab:othersearches}
\end{deluxetable}

\begin{deluxetable}{ccccccc}
\tabletypesize{\scriptsize}
\tablecaption{Known \lya\ Nebulae in the NDWFS Bo\"otes Field}
\tablewidth{0pt}
\tablehead{
\colhead{Source} & \colhead{Survey\tablenotemark{a}} & \colhead{Redshift} & \colhead{\bw\ Isophotal Area\tablenotemark{b}} & \colhead{\bw\ Surface Brightness\tablenotemark{c}} & \colhead{Notes\tablenotemark{d}} \\
 &  &  & (arcsec$^{2}$) & (mag arcsec$^{-2}$) &  \\
 }
\startdata
    LABd05 &                             Dey et al. (2005) &      2.7 &  54.4 &  27.0 &                 Recovered   \\
        P1 &                      Prescott et al. - Subaru &      2.7 &  27.6 &  27.3 &                 Too Faint   \\
        P2 &                      Prescott et al. - Subaru &      2.6 &  71.0 &  26.9 &                 Core-Halo   \\
        P3 &                      Prescott et al. - Mayall &      1.9 &  25.4 &  26.9 &                 Core-Halo   \\
        Y1 &                            Yang et al. (2009) &      2.3 &  20.4 &  26.7 &                 Core-Halo   \\
        Y2 &                            Yang et al. (2009) &      2.3 &  17.0 &  26.8 &                 Core-Halo   \\
        Y3 &                            Yang et al. (2009) &      2.3 &  36.6 &  25.9 &                 Core-Halo   \\
        Y4 &                            Yang et al. (2009) &      2.3 &  37.5 &  26.6 &                 Core-Halo   \\
\enddata
\tablenotetext{a}{Spectroscopically confirmed \lya\ nebulae selected from a Subaru intermediate-band survey (Prescott et al. 2012, in preparation) and a Mayall intermediate-band survey (Prescott et al. 2012, in preparation) are listed along with \lya\ nebulae from the narrow-band survey by Yang et al. (2009).}
\tablenotetext{b}{Isophotal area determined using SourceExtractor and a detection threshold set to the median 1$\sigma$ surface brightness threshold of the NDWFS \bw\ imaging (Section~\ref{sec:comparison}).}
\tablenotetext{c}{Mean surface brightness within the isophotal area determined using SourceExtractor (Section~\ref{sec:comparison}).}
\tablenotetext{d}{``Too Faint" indicates that the diffuse emission was too faint to be selected by our survey.  ``Core-Halo" describes sources with a central compact continuum source surrounded by a \lya\ halo; these sources are partially or completely removed from the images during pipeline steps designed to remove bright stars and compact galaxies and are not selected by our survey.}
\label{tab:otherblob}
\end{deluxetable}

\begin{figure}
\center
\includegraphics[angle=0,width=6.5in]{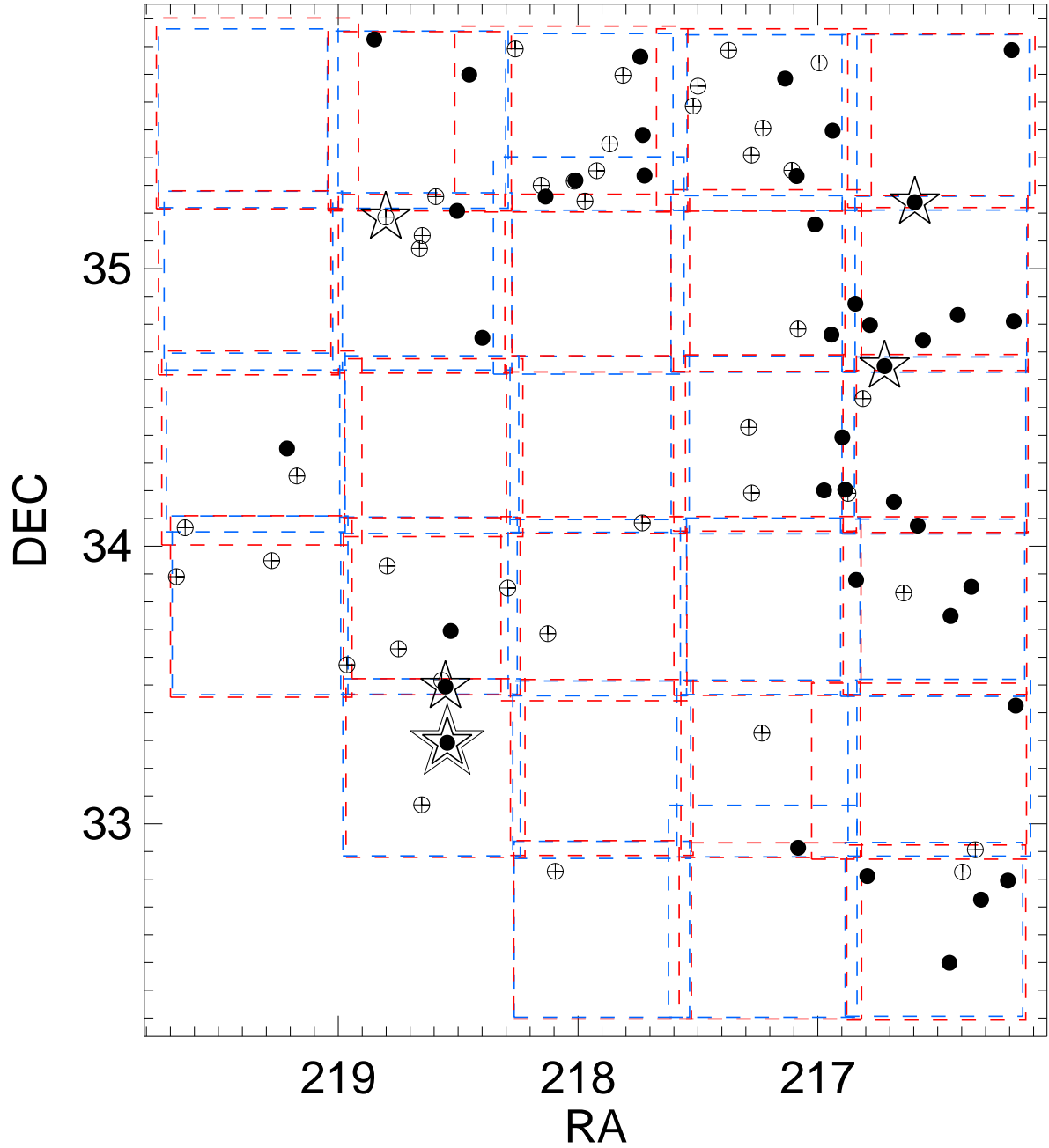}
\caption{Sky distribution of priority 1 (filled circles) and priority 2 (crossed open circles) \lya\ nebula candidates within 
the NDWFS Bo\"otes field.  The individual NDWFS tiles are shown as blue (\bw\ imaging) and red ($R$-band imaging) 
dashed rectangles.  Confirmed \lya\ nebulae are shown as stars (Paper~II), including LABd05 \citep[double star;][]{dey05}.  
}
\label{fig:footprint}
\end{figure}

\clearpage 

\appendix
\section{Postage Stamps}
\label{appendixA}

Postage stamp images of the \numcand\ \lya\ nebula candidates 
from our broad-band survey of the NDWFS Bo\"otes field.  
The \bw\ and $R$-band images shown were used for morphological and color selection 
during the survey; the $I$-band images are shown for reference.

\begin{figure}
\vspace{0.5in}
\center
\plottwo{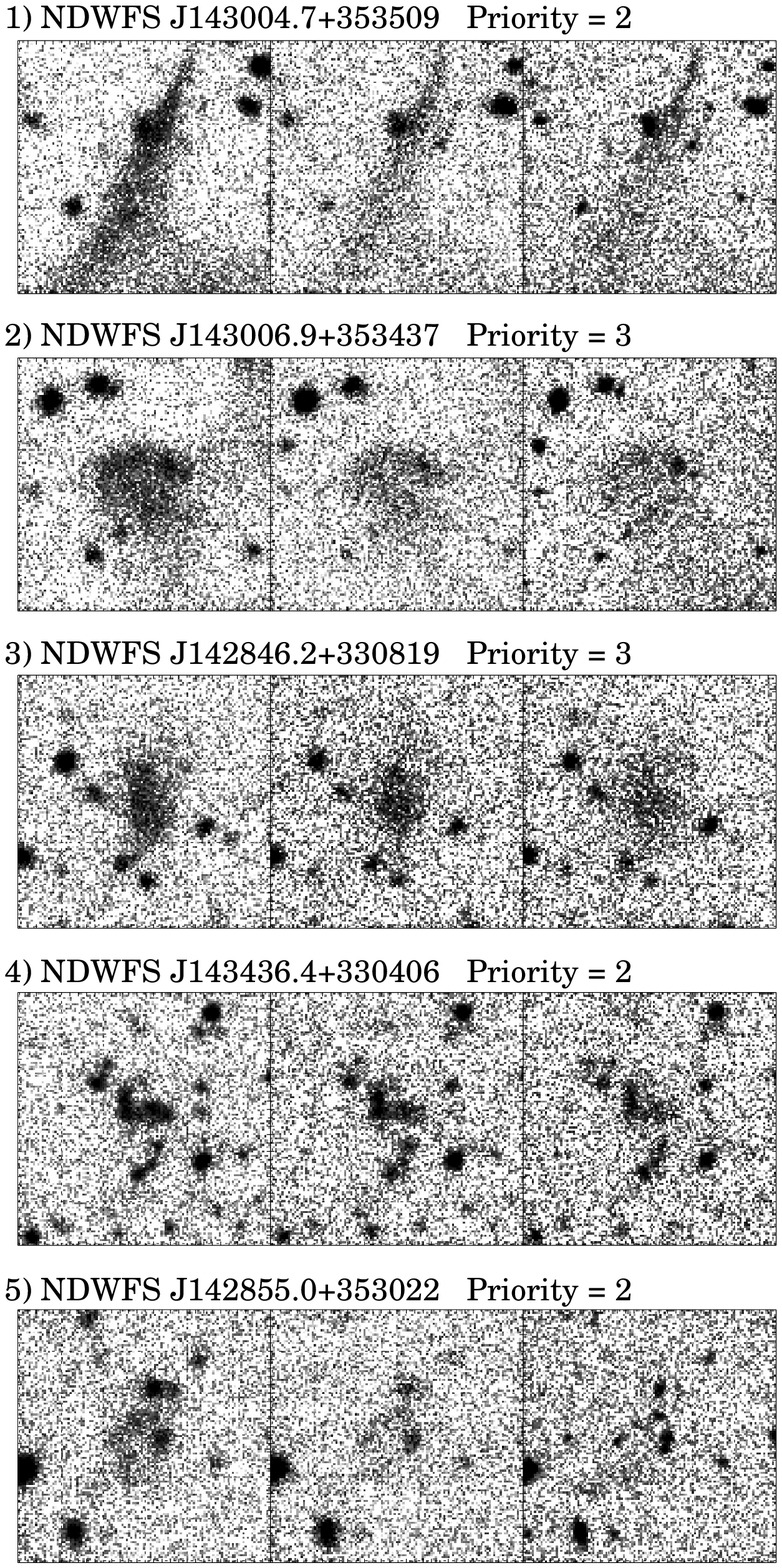}{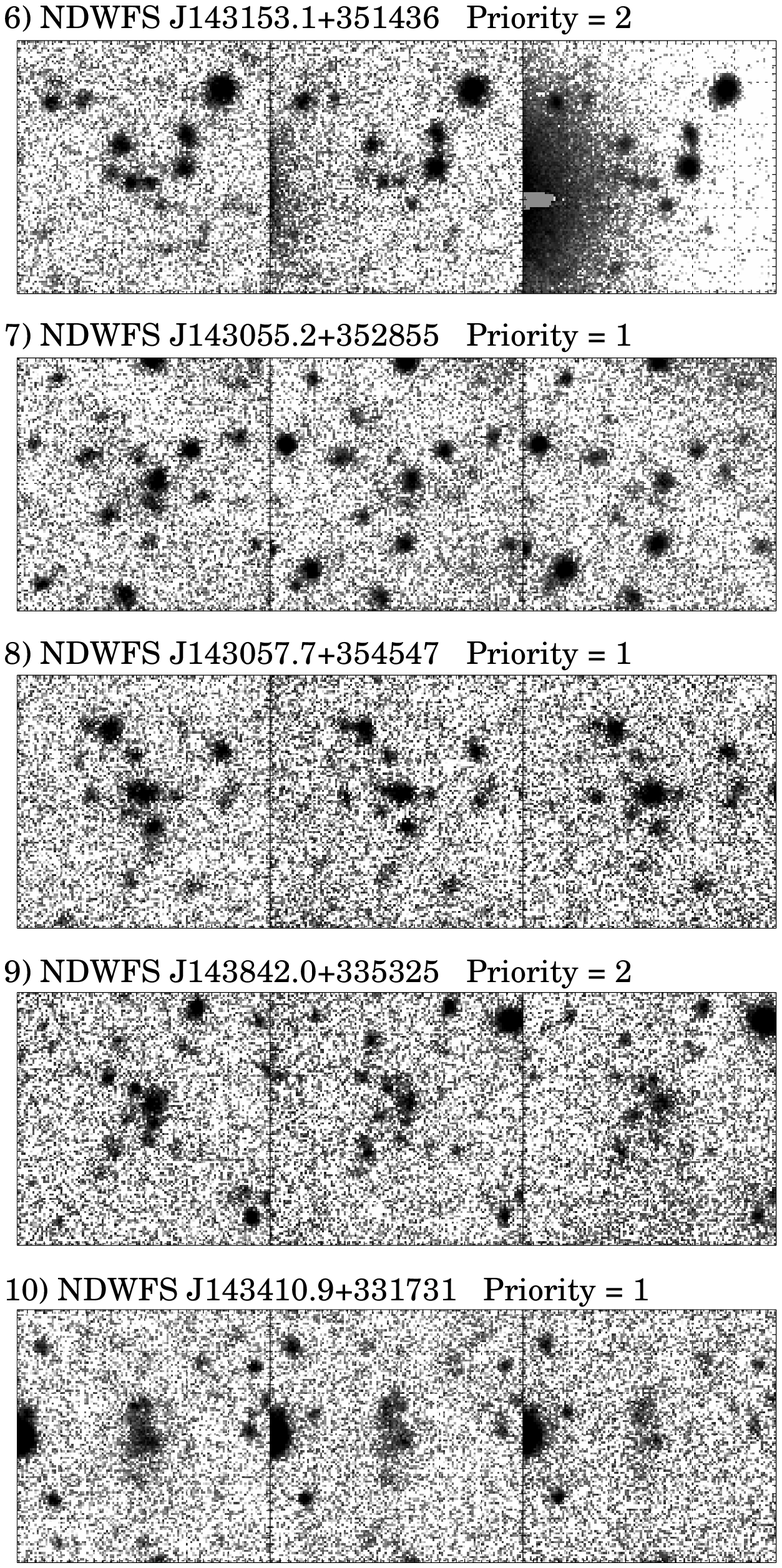}
\caption{
Postage stamp images of the \lya\ nebula candidates in the \bw, $R$, and $I$ bands, respectively.  
Images are displayed using a log scaling within 30\arcsec$\times$30\arcsec\ boxes, 
and are labeled with the candidate number, name, and priority listed in Table~\ref{tab:select}.  
}
\label{fig:appendixA_1}
\end{figure}

\begin{figure}
\vspace{0.5in}
\center
\plottwo{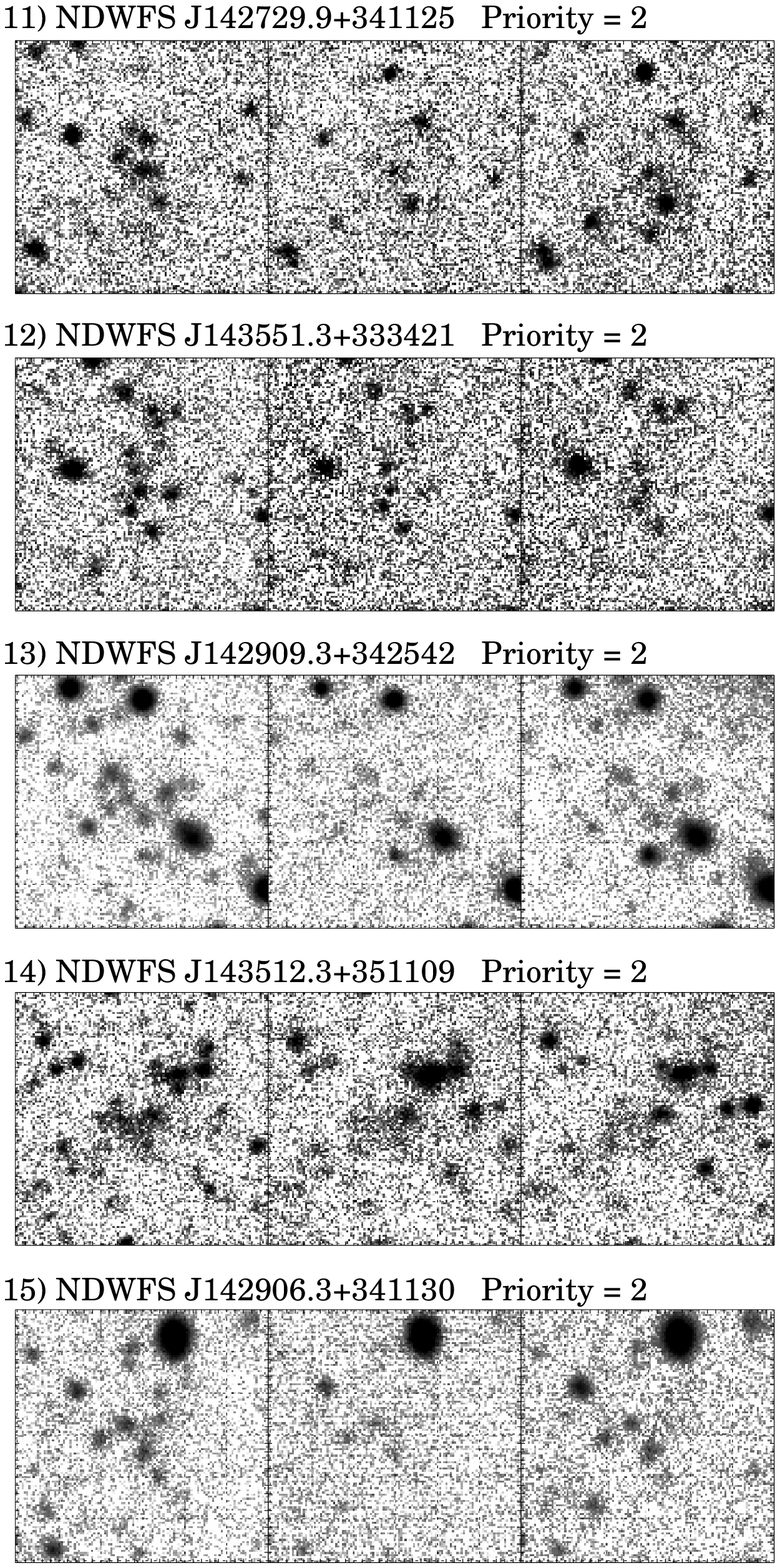}{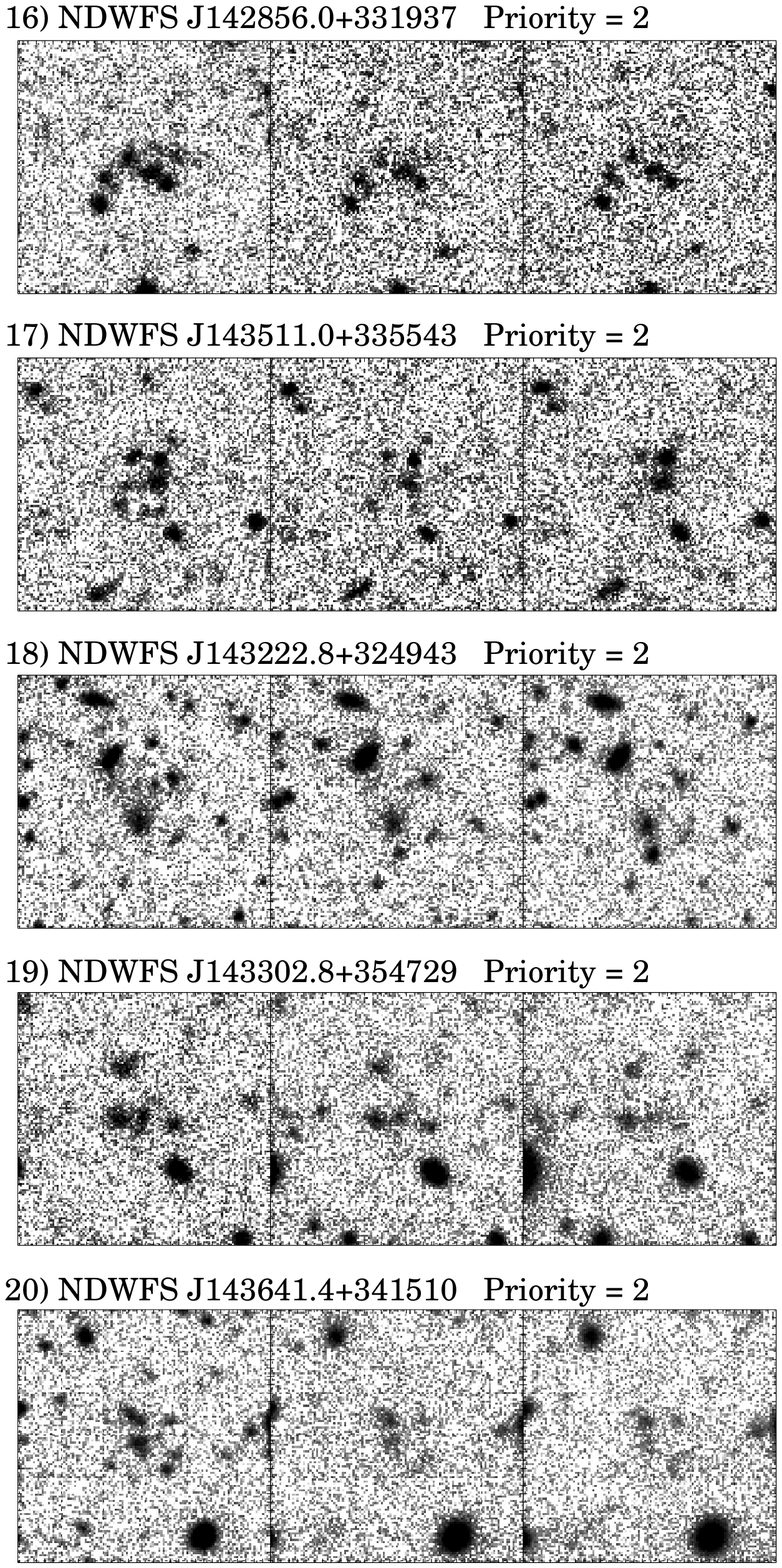}
\caption{ 
Postage stamp images of the \lya\ nebula candidates in the \bw, $R$, and $I$ bands, respectively.  
Images are displayed using a log scaling within 30\arcsec$\times$30\arcsec\ boxes, 
and are labeled with the candidate number, name, and priority listed in Table~\ref{tab:select}.  
}
\label{fig:appendixA_2}
\end{figure}

\begin{figure}
\vspace{0.5in}
\center
\plottwo{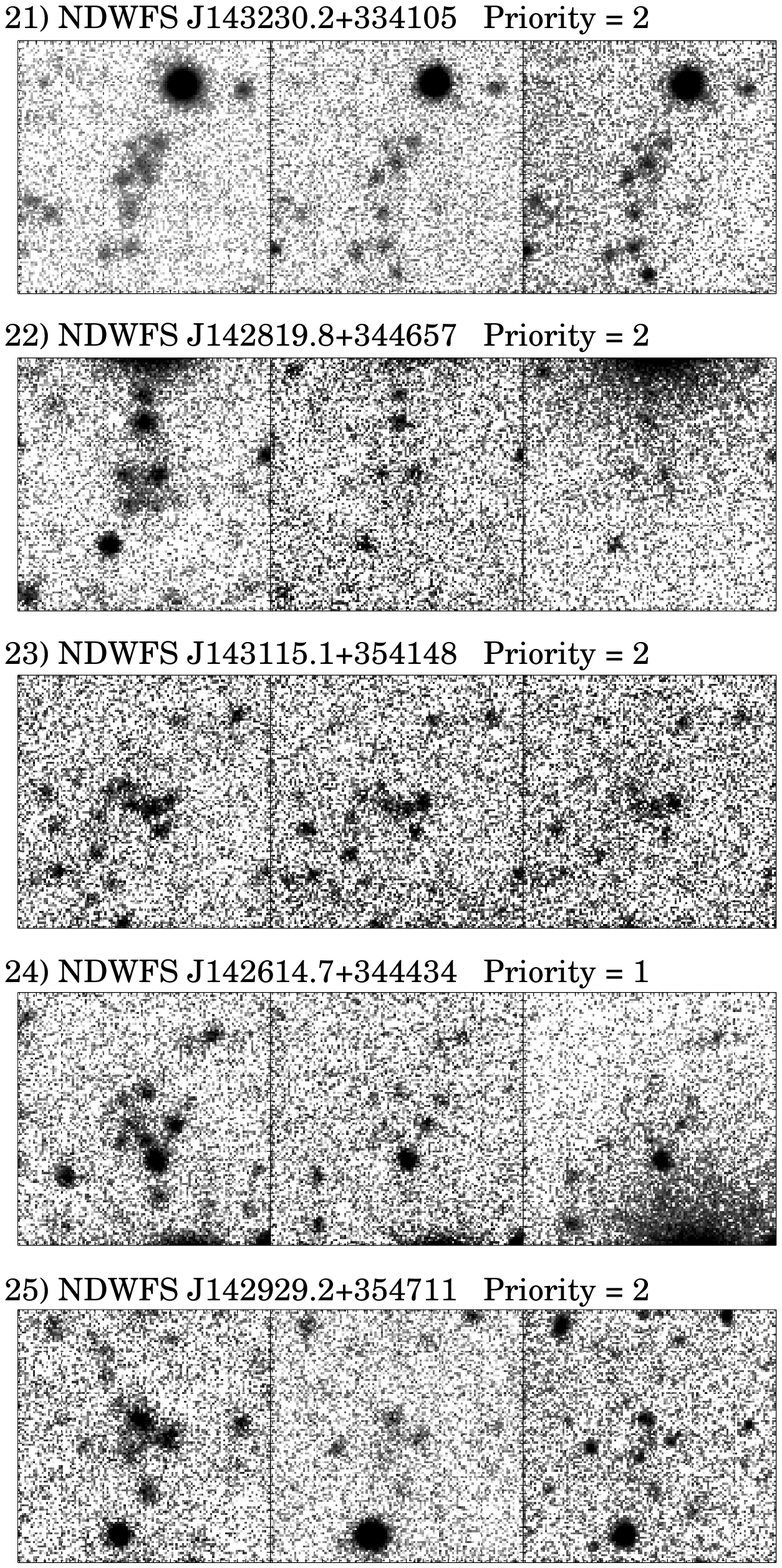}{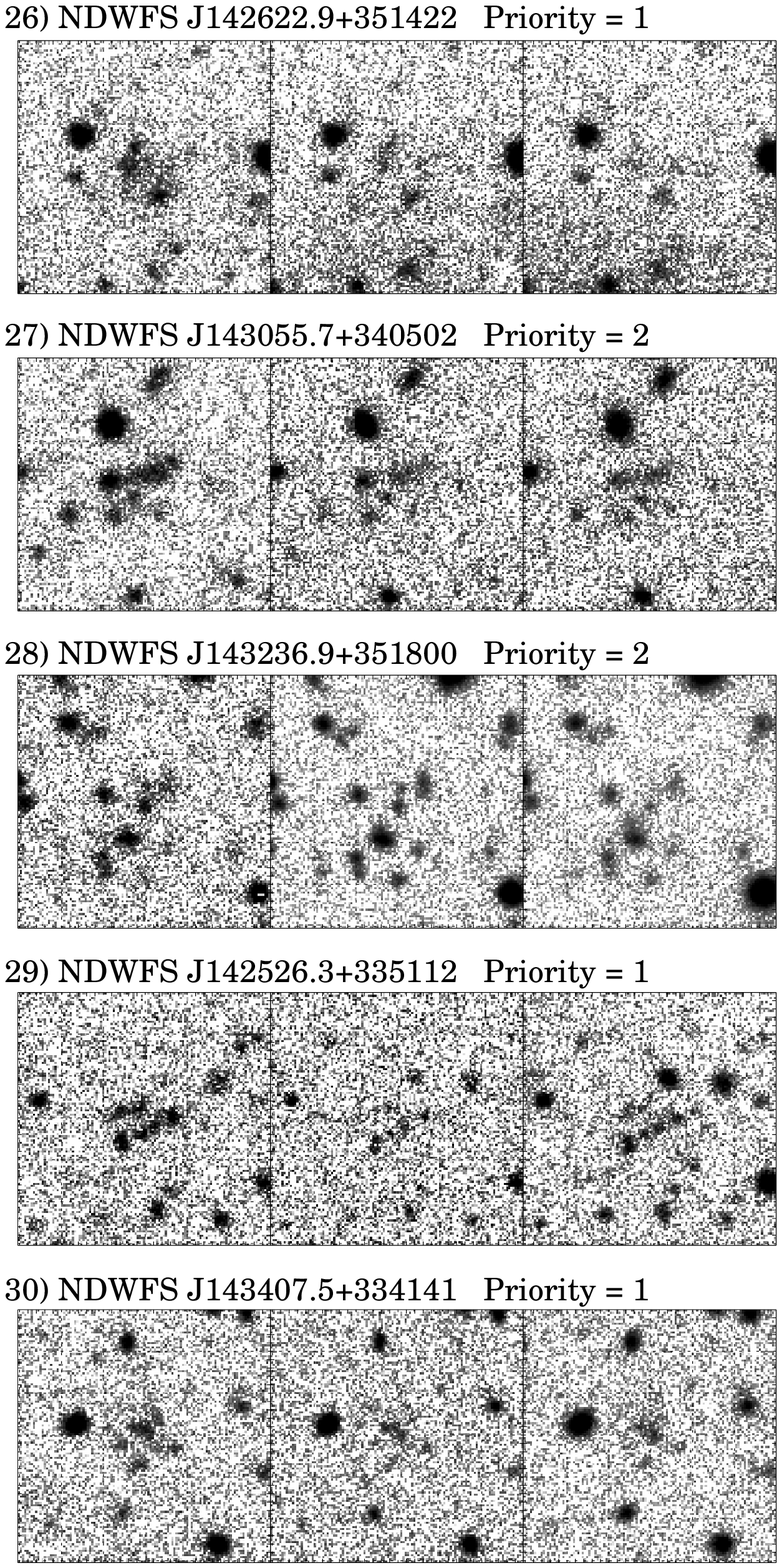}
\caption{ 
Postage stamp images of the \lya\ nebula candidates in the \bw, $R$, and $I$ bands, respectively.  
Images are displayed using a log scaling within 30\arcsec$\times$30\arcsec\ boxes, 
and are labeled with the candidate number, name, and priority listed in Table~\ref{tab:select}.  
}
\label{fig:appendixA_3}
\end{figure}

\begin{figure}
\vspace{0.5in}
\center
\plottwo{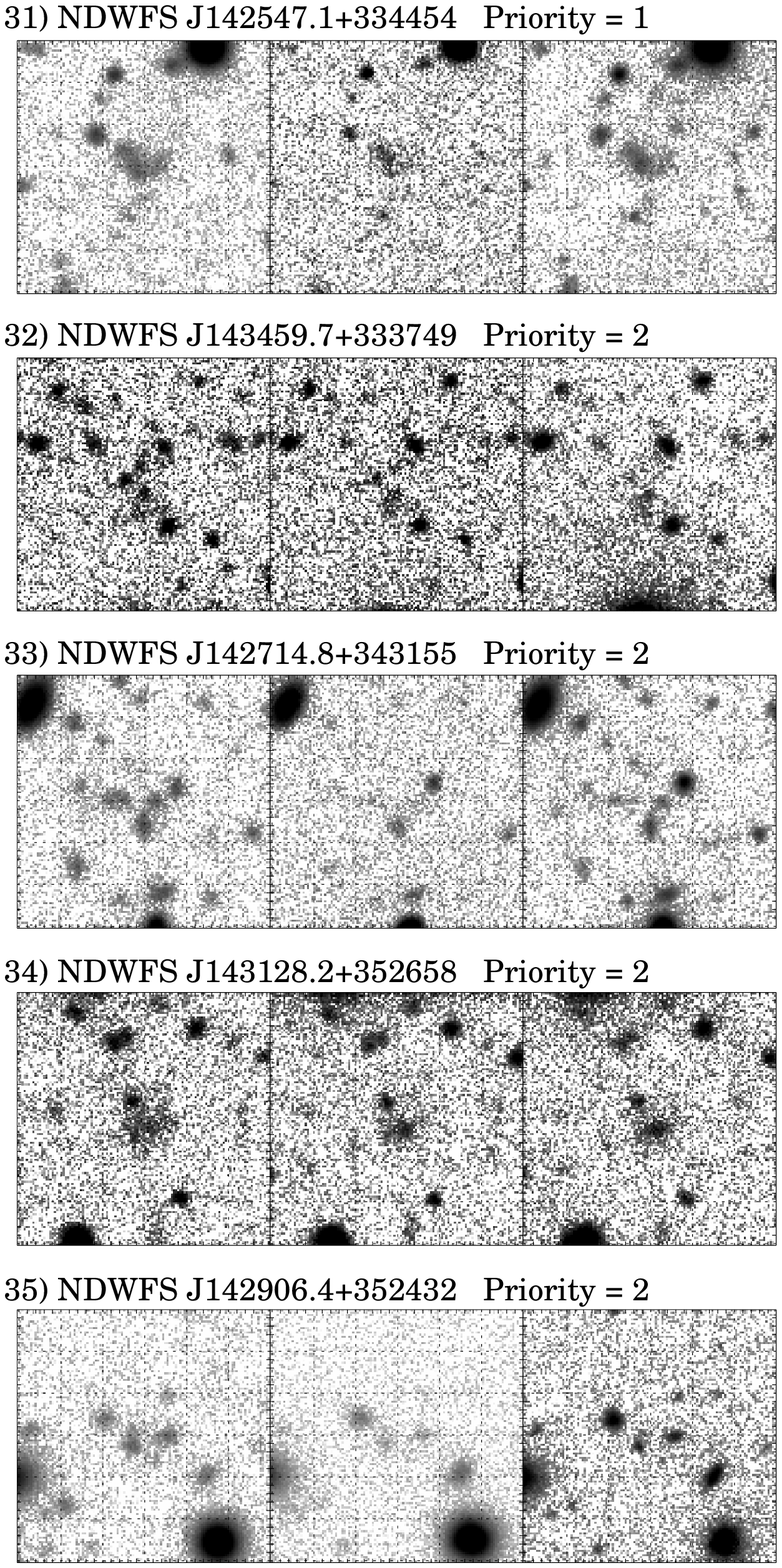}{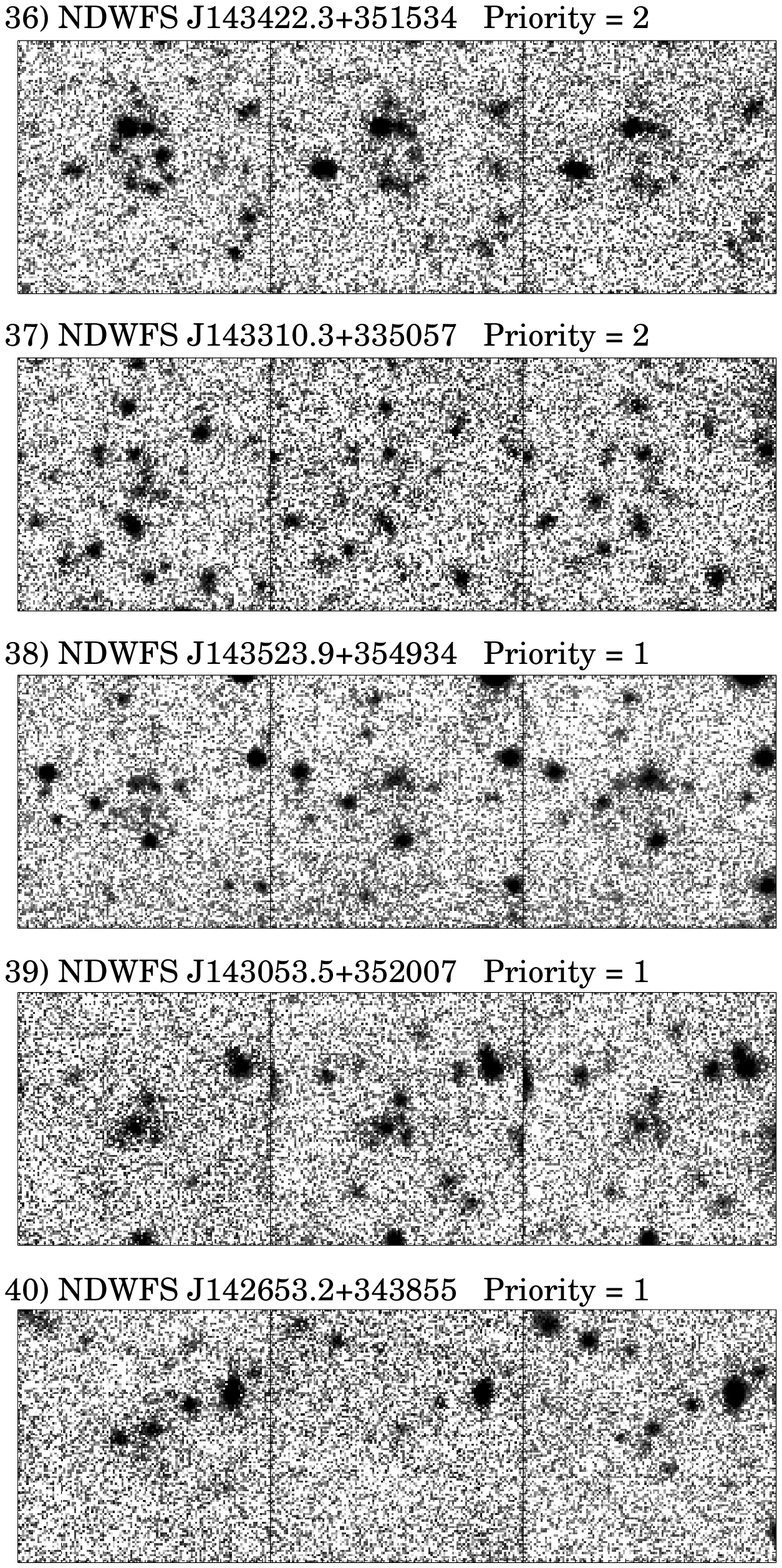}
\caption{ 
Postage stamp images of the \lya\ nebula candidates in the \bw, $R$, and $I$ bands, respectively.  
Images are displayed using a log scaling within 30\arcsec$\times$30\arcsec\ boxes, 
and are labeled with the candidate number, name, and priority listed in Table~\ref{tab:select}.  
}
\label{fig:appendixA_4}
\end{figure}

\begin{figure}
\vspace{0.5in}
\center
\plottwo{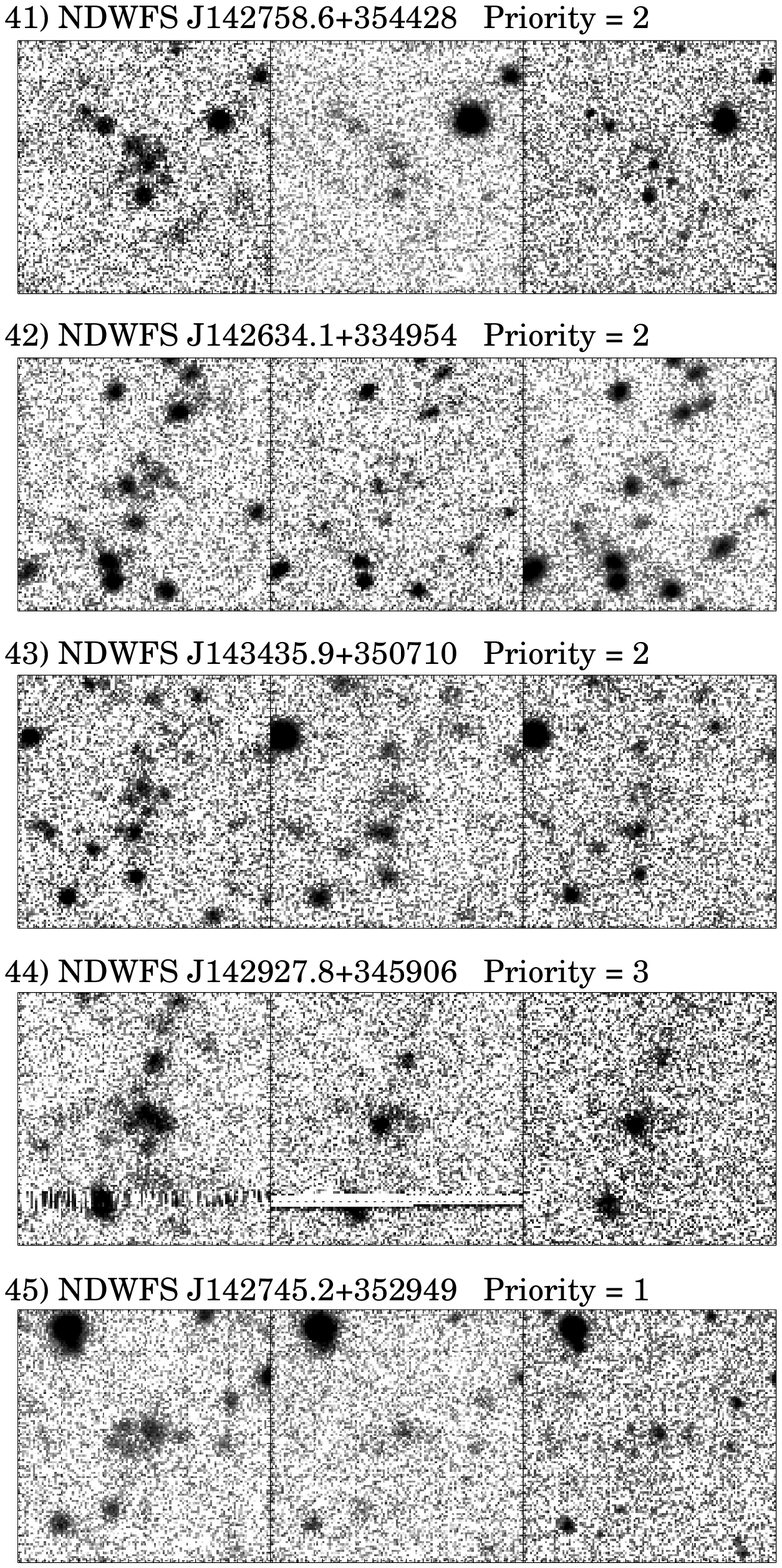}{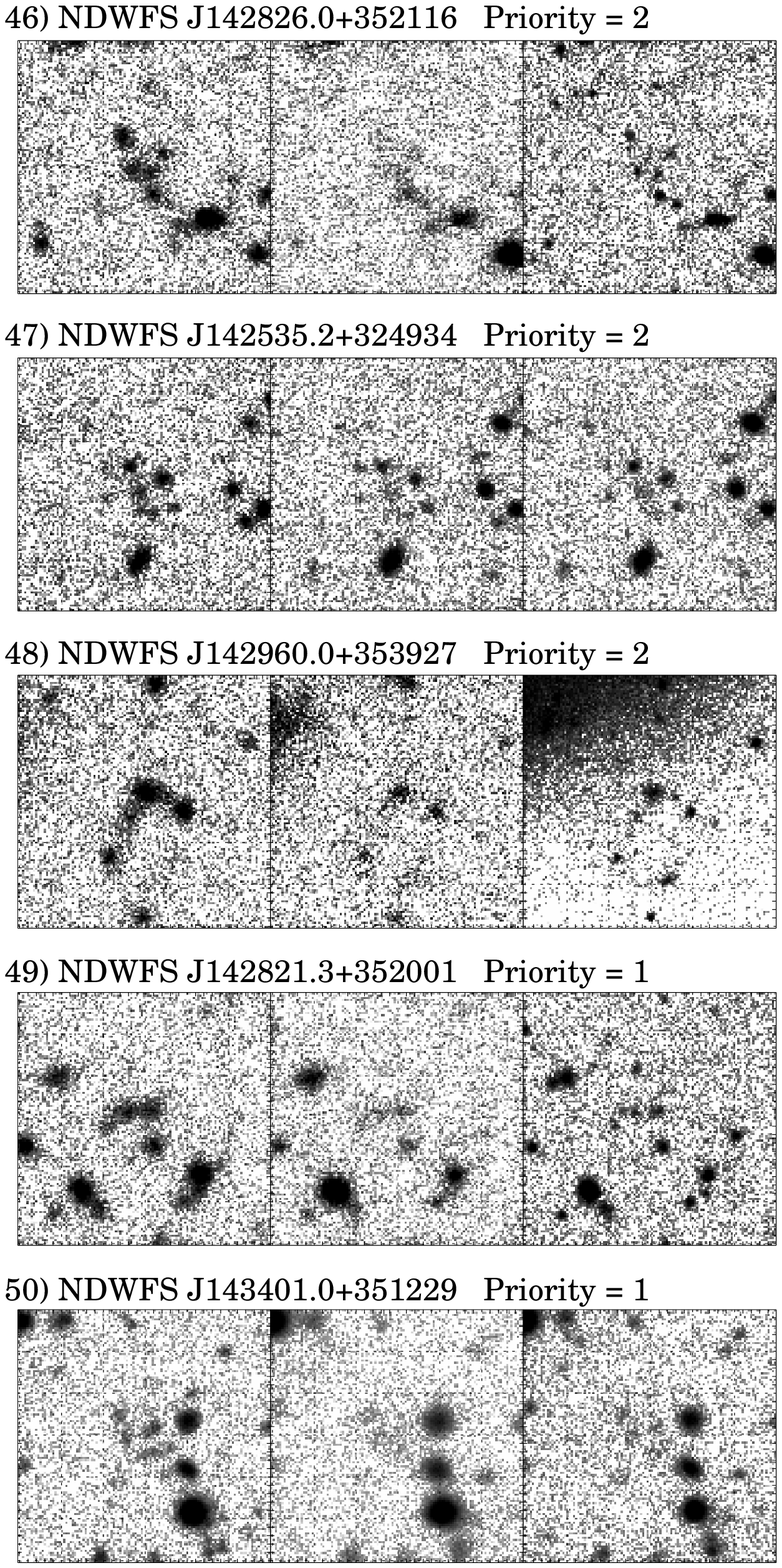}
\caption{ 
Postage stamp images of the \lya\ nebula candidates in the \bw, $R$, and $I$ bands, respectively.  
Images are displayed using a log scaling within 30\arcsec$\times$30\arcsec\ boxes, 
and are labeled with the candidate number, name, and priority listed in Table~\ref{tab:select}.  
}
\label{fig:appendixA_5}
\end{figure}

\begin{figure}
\vspace{0.5in}
\center
\plottwo{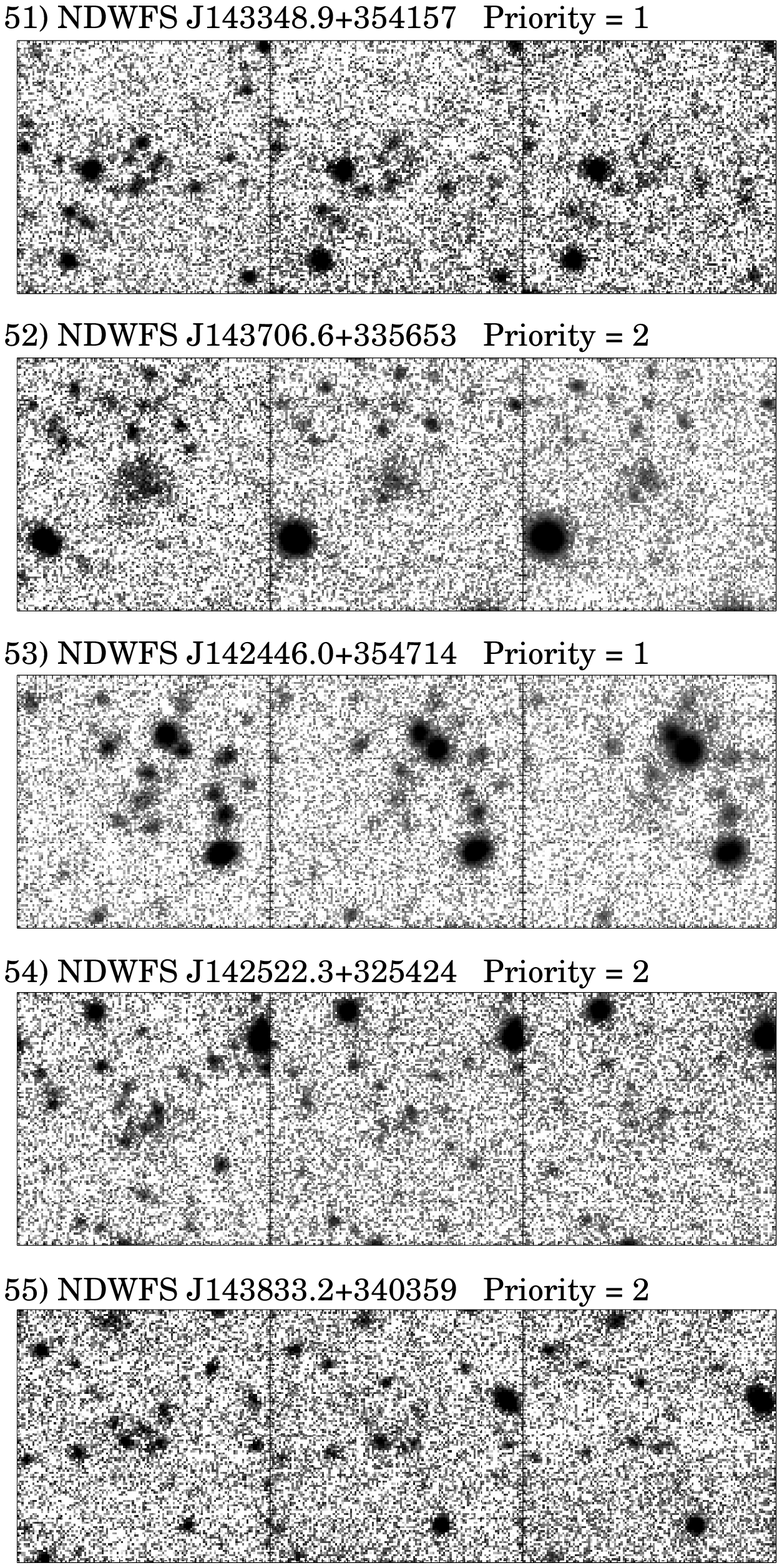}{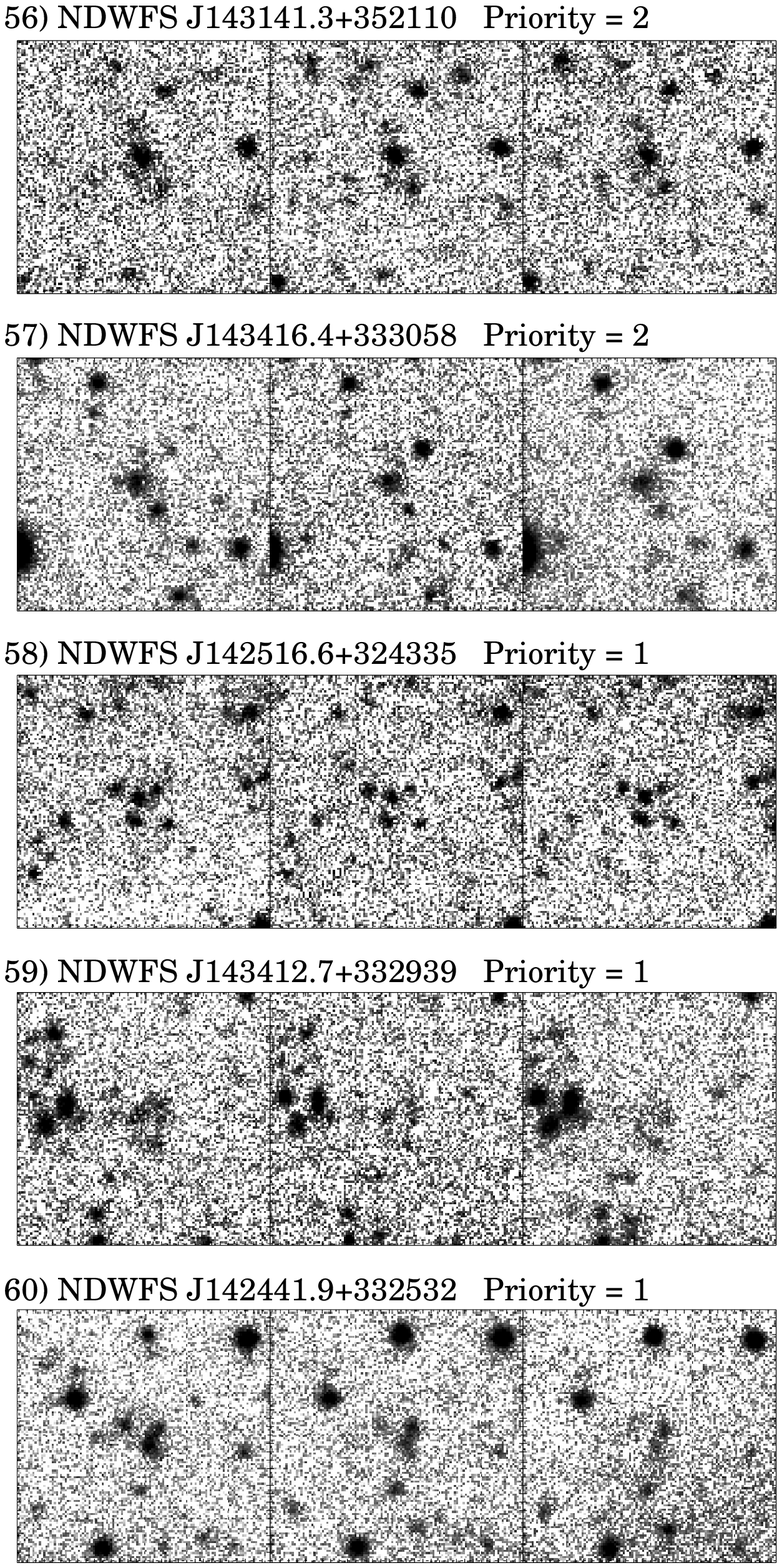}
\caption{ 
Postage stamp images of the \lya\ nebula candidates in the \bw, $R$, and $I$ bands, respectively.  
Images are displayed using a log scaling within 30\arcsec$\times$30\arcsec\ boxes, 
and are labeled with the candidate number, name, and priority listed in Table~\ref{tab:select}.  
}
\label{fig:appendixA_6}
\end{figure}

\begin{figure}
\vspace{0.5in}
\center
\plottwo{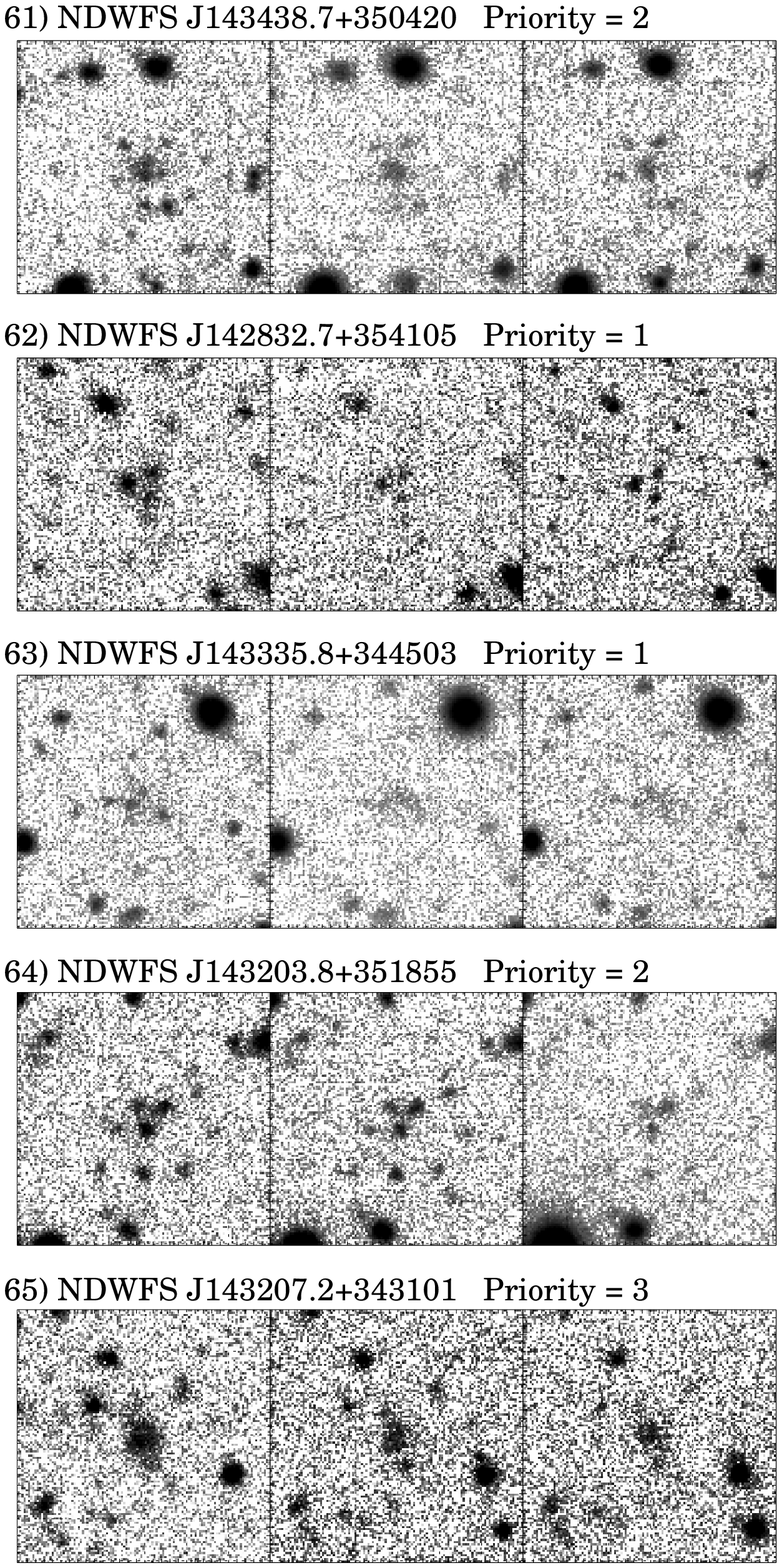}{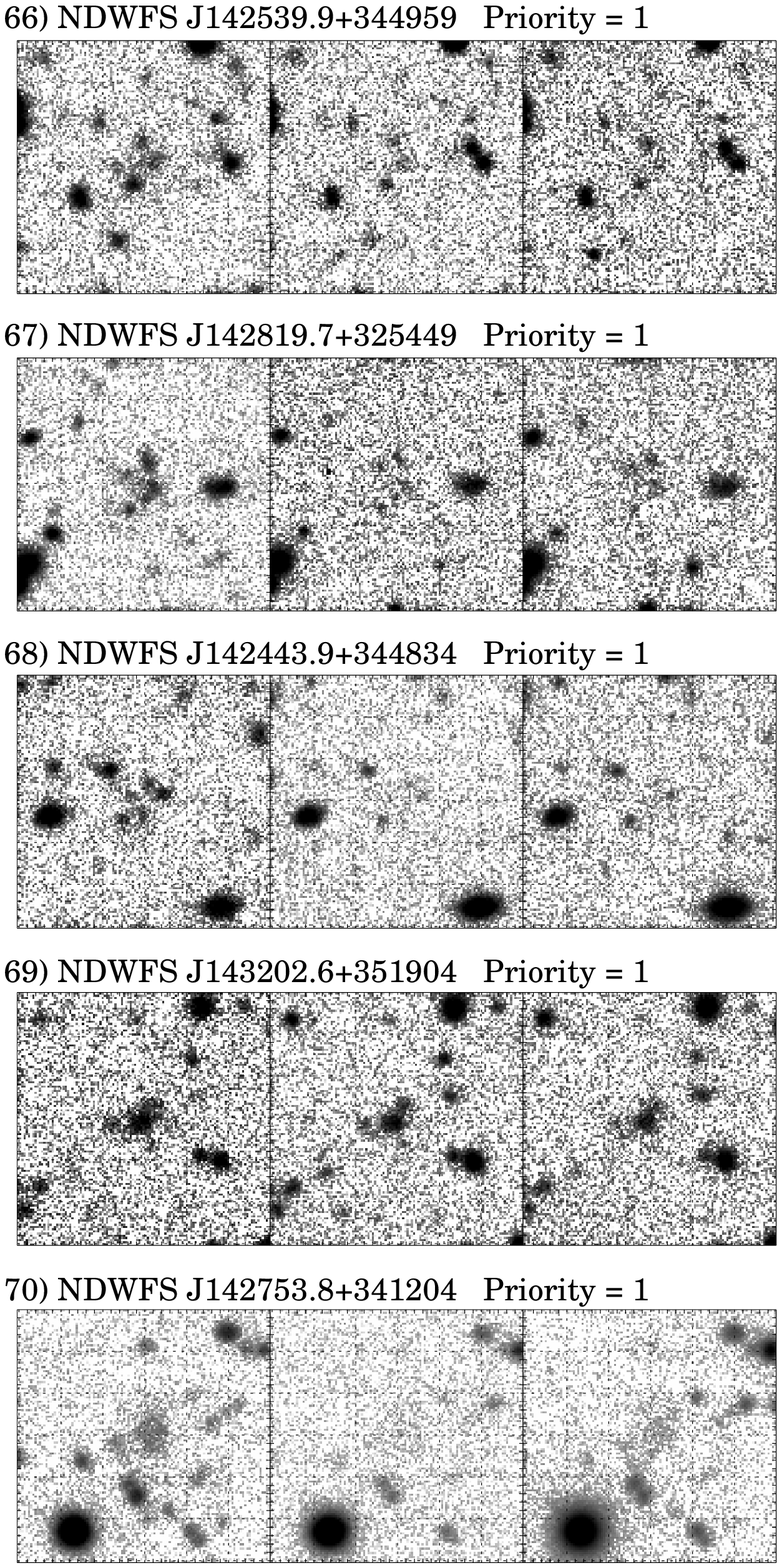}
\caption{ 
Postage stamp images of the \lya\ nebula candidates in the \bw, $R$, and $I$ bands, respectively.  
Images are displayed using a log scaling within 30\arcsec$\times$30\arcsec\ boxes, 
and are labeled with the candidate number, name, and priority listed in Table~\ref{tab:select}.  
}
\label{fig:appendixA_7}
\end{figure}

\begin{figure}
\vspace{0.5in}
\center
\plottwo{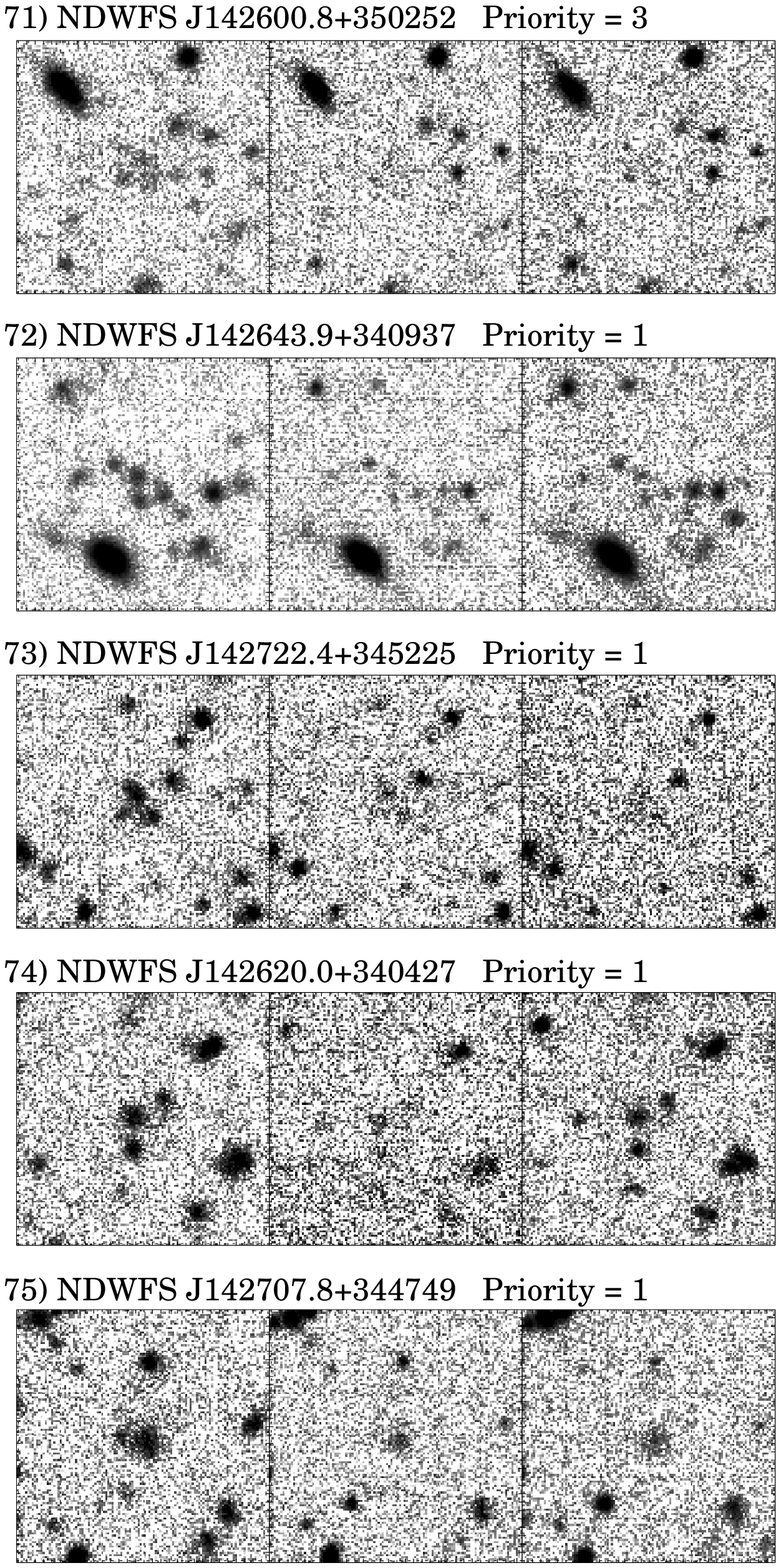}{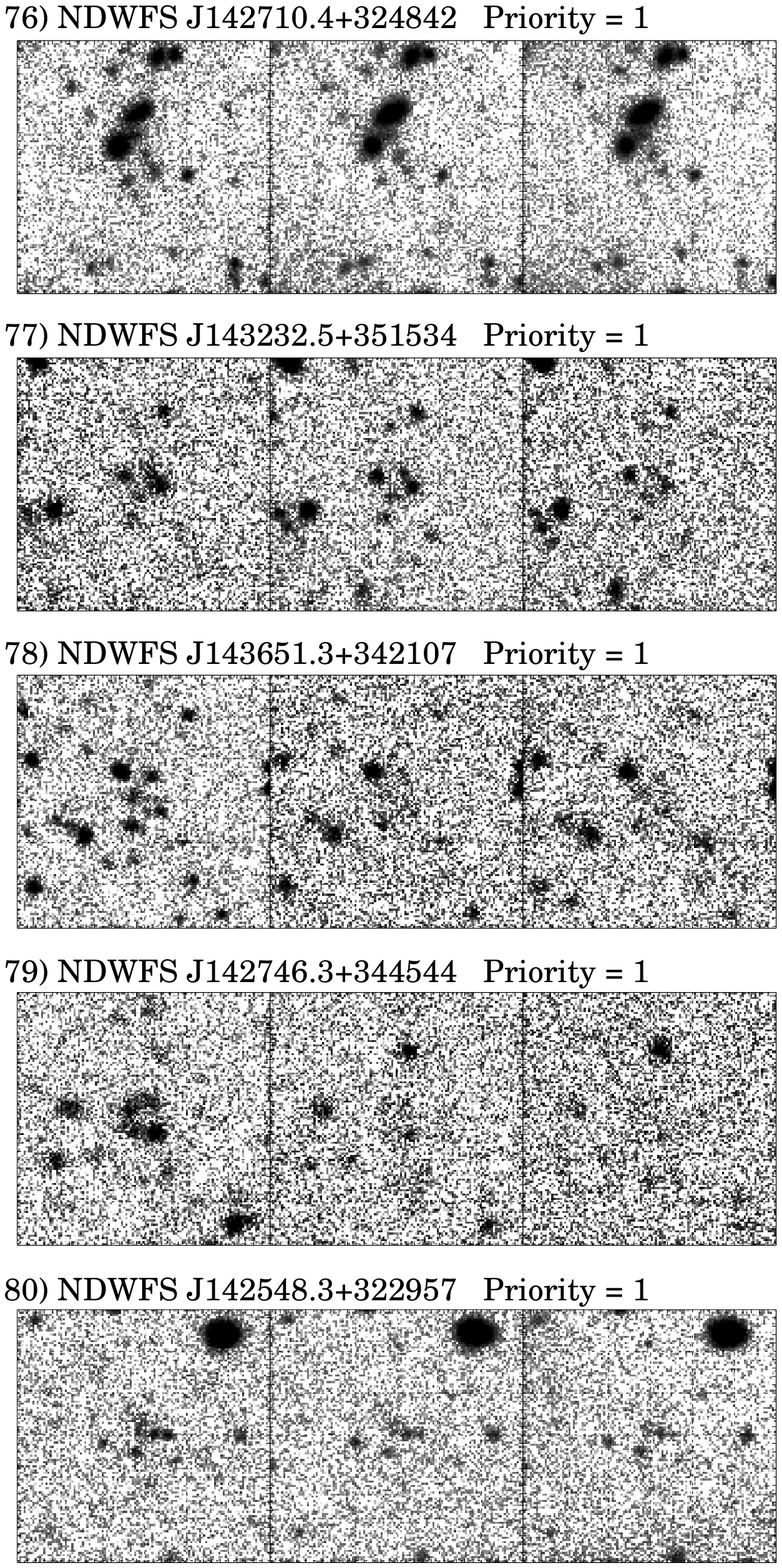}
\caption{ 
Postage stamp images of the \lya\ nebula candidates in the \bw, $R$, and $I$ bands, respectively.  
Images are displayed using a log scaling within 30\arcsec$\times$30\arcsec\ boxes, 
and are labeled with the candidate number, name, and priority listed in Table~\ref{tab:select}.  
}
\label{fig:appendixA_8}
\end{figure}

\begin{figure}
\vspace{0.5in}
\center
\includegraphics[angle=0,width=3in]{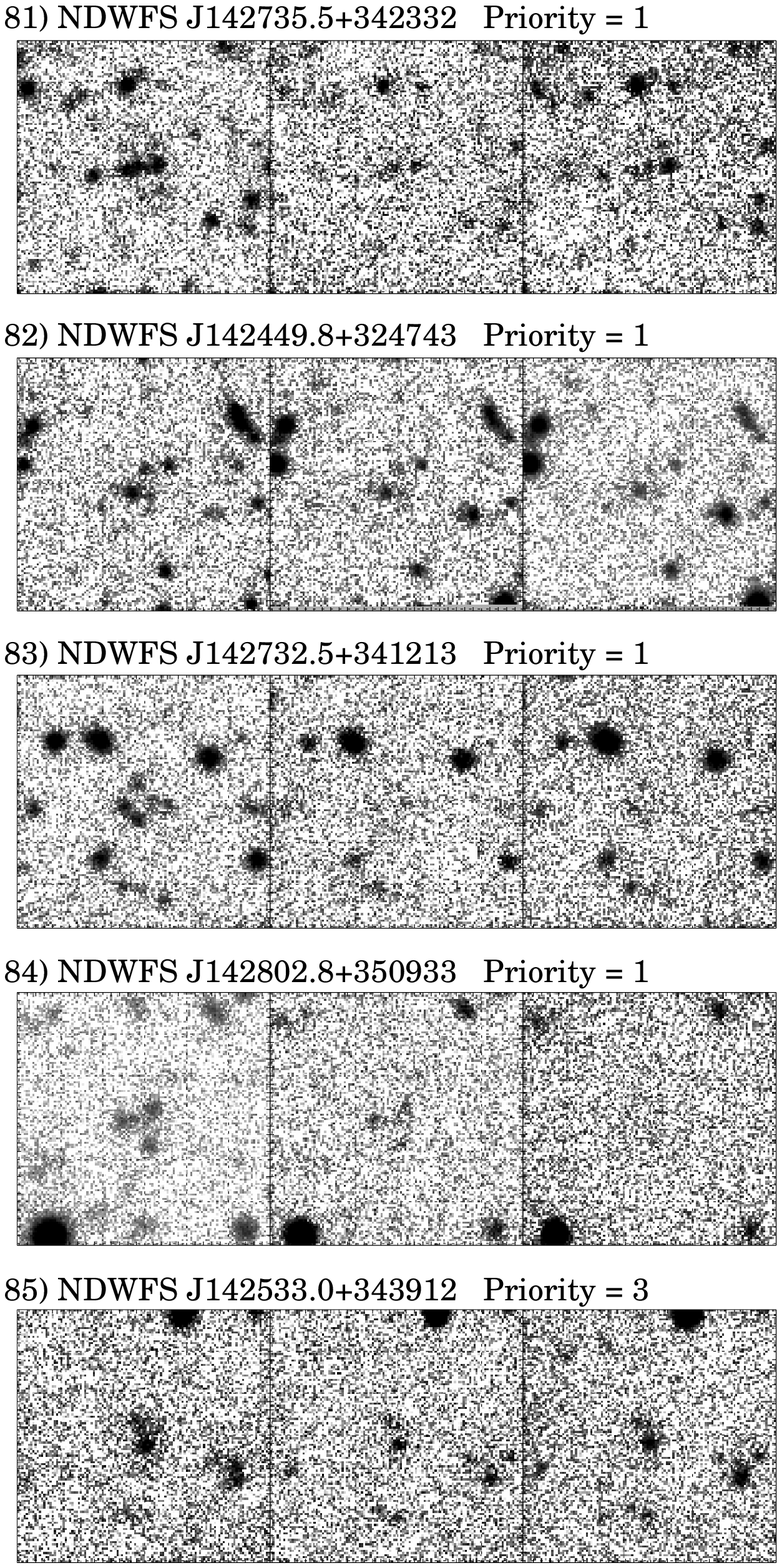}
\caption{ 
Postage stamp images of the \lya\ nebula candidates in the \bw, $R$, and $I$ bands, respectively.  
Images are displayed using a log scaling within 30\arcsec$\times$30\arcsec\ boxes, 
and are labeled with the candidate number, name, and priority listed in Table~\ref{tab:select}.  
}
\label{fig:appendixA_9}
\end{figure}

\end{document}